\documentclass[journal=jctcce,manuscript=article,layout=twocolumn]{achemso}
\setkeys{acs}{articletitle = true, chaptertitle = true, doi = true}

\usepackage[version=3]{mhchem} 
\usepackage[T1]{fontenc}       
\usepackage{mathptmx}

\makeatletter
\let\l@addto@macro\relax
\makeatother
\usepackage[fontsize=9pt]{scrextend}

\usepackage{graphicx}  
\usepackage{bm}        
\usepackage{bbm}       
\usepackage{amssymb}   
\usepackage{amsmath}   
\usepackage{physics}   
\usepackage{textgreek} 
\usepackage{subfigure} 
\usepackage{multirow}  
\usepackage{booktabs}  
\usepackage{hyperref}  
\hypersetup{colorlinks=true, linkcolor=black, urlcolor=black, citecolor=black, plainpages=false,
bookmarksnumbered=true}
\usepackage{xurl}      
\usepackage{siunitx}   
\usepackage{xcolor}    
\usepackage[utf8]{inputenc} 
\author{Christof Holzer}
\affiliation{Institute of Theoretical Solid State Physics, Karlsruhe
Institute of Technology (KIT), Wolfgang-Gaede-Stra\ss{}e 1, 76131 Karlsruhe,
Germany}
\email{christof.holzer@kit.edu}

\author{Yannick J. Franzke}
\affiliation{Otto Schott Institute of Materials Research,
Friedrich Schiller University Jena, L{\"o}bdergraben 32, 07743 Jena, Germany}
\email{yannick.franzke@uni-jena.de}

\title{A General and Transferable Local Hybrid Functional for Electronic Structure Theory and Many-Fermion Approaches}

\begin{document}



\begin{abstract}
Density functional theory has become the workhorse of quantum physics, chemistry,
and materials science. Within these fields, a broad range of applications needs 
to be covered. These applications range from solids to molecular systems, 
from organic to inorganic chemistry, or even from electrons to other 
fermions such as protons or muons.
This is emphasized by the plethora of density functional
approximations that have been developed for various cases.
In this work, a new local hybrid exchange-correlation density functional is
constructed from first principles, promoting generality and transferability.
We show that constraint satisfaction can be achieved even for admixtures with 
full exact exchange, without sacrificing accuracy.
The performance of the new functional for electronic structure theory is
assessed for thermochemical properties, excitation energies, M\"ossbauer
isomer shifts, NMR spin--spin coupling constants, NMR shieldings and shifts,
magnetizabilities, as well as EPR hyperfine coupling constants.
Here, the new density functional shows excellent performance throughout all tests
and is numerically robust only requiring small grids for converged results.
Additionally, the functional can be easily generalized to arbitrary fermions
as shown for electron-proton correlation energies.
Therefore, we outline that density functionals generated in this way are 
general purpose tools for quantum mechanical studies.
\end{abstract}

\section{Introduction}
\label{sec:introduction}
Density functional theory (DFT) is very likely the most commonly applied
computational method for electronic structure theory in physics, chemistry,
materials science, and related fields.
This success stems from a favorable cost-accuracy ratio making DFT applicable
to very large systems with good accuracy. \cite{Csonka.Perdew.ea:Global.2010,
Burke:Perspective.2012, Becke:Perspective.2014, Mardirossian.Head-Gordon:Thirty.2017}
Both ``pure'' or semilocal DFT as well as hybrid DFT methods can be applied
in a black-box fashion and are computationally cheaper than all
wavefunction-based methods including exchange and correlation.
The prize to pay is a dependence of the
results on the underlying density functional approximation (DFA),
which is commonly classified with Jacob's ladder. \cite{Perdew.Schmidt:Jacobs.2001}
These DFAs are either designed with large molecular data sets and many
fitting parameters or by considering theoretical constraints and data
of the noble gases in a more \textit{ab initio} fashion.
\cite{Burke:Perspective.2012, Becke:Perspective.2014}
Prominent examples of the \textit{ab initio} way are the
modern meta-generalized gradient approximations (meta-GGAs)
developed by the groups of Perdew, Tao, and Sun.
\cite{Tao.Mo:Accurate.2016, Sun.Ruzsinszky.ea:Strongly.2015,
Furness.Kaplan.ea:Accurate.2020, Furness.Kaplan.ea:Correction.2020}
Designing functionals from first principles may yield results
inferior to highly parameterized DFAs for the corresponding test
or data set. However, it comes with more generality and physical insight.
\cite{Medvedev.Bushmarinov.ea:Density.2017, Becke:Density-Functional.2022}
Of course, combinations of the two design philosophies, i.e.\ taking the
best of both, are possible. \cite{Becke:Density-functional.a2022}

Of particular interest for the development of a general functional
is the self-interaction error. In this regard,
local hybrid functionals \cite{Jaramillo.Scuseria.ea:Local.2003} (LHFs)
offer an increased flexibility over the more common global
\cite{Becke:Density-functional.1993, Stephens.Devlin.ea:Ab.1994}
and range-separated hybrid functionals, \cite{Gill.Adamson.ea:Coulomb-attenuated.1996,
Leininger.Stoll.ea:Combining.1997, Iikura.Tsuneda.ea:Long.2001,
Yanai.Tew.ea:new.2004} as LHFs use a fully position dependent amount of
exact exchange. Therefore, LHFs allow to switch from 0\% exact exchange
to 100\% exact exchange, which is advantageous for strongly localized
fermions such as protons. \cite{Holzer.Franzke:Beyond.2024}
The corresponding local mixing functions (LMFs) are,
for instance, based on the iso-orbital indicator \cite{Jaramillo.Scuseria.ea:Local.2003}
(t-LMF) or the correlation length (z-LMF). \cite{Johnson:Local-hybrid.2014}
Within the last 20 years, much effort was put into
constructing LMFs and exchange contributions.
\cite{Johnson:Local-hybrid.2014, Schmidt.Kraisler.ea:self-interaction-free.2014,
Arbuznikov.Kaupp:Local.2007, Arbuznikov.Kaupp:Towards.2014, Haasler.Maier.ea:Local.2020,
Tao.Staroverov.ea:Exact-exchange.2008, Holzer.Franzke:Local.2022,
Silva.Corminboeuf:Local.2015, Janesko.Scuseria:Local.2007,
Janesko.Krukau.ea:Self-consistent.2008, Janesko.Scuseria:Parameterized.2008,
Maier.Arbuznikov.ea:Local.2019, Janesko:Replacing.2021, Grotjahn:Learning.2023}
At the same time, efficient implementations and applications for a wide range
of properties from the ground state 
\cite{Plessow.Weigend:Seminumerical.2012, Laqua.Kussmann.ea:Efficient.2018,
Klawohn.Bahmann.ea:Implementation.2016, Holzer.Franzke.ea:Assessing.2021,
Jimenez-Hoyos.Janesko.ea:Assessment.2009, Liu.Proynov.ea:Comparison.2012}
to excited states
\cite{Maier.Bahmann.ea:Efficient.2015, Grotjahn.Furche.ea:Development.2019,
Kehry.Franzke.ea:Quasirelativistic.2020, Holzer:improved.2020,
Zerulla.Krstic.ea:Multi-Scale.2022, Zerulla.Venkitakrishnan.ea:T-Matrix.2023,
Zerulla.Li.ea:Exploring.2023, Mueller.Perdana.ea:Modeling.2022,
Rai.Balzer.ea:Hot.2023, Rai.Gerhard.ea:Activating.2023}
and magnetic properties \cite{Schattenberg.Reiter.ea:Efficient.2020,
Mack.Schattenberg.ea:Nuclear.2020, Franzke.Mack.ea:NMR.2021, 
Franzke.Holzer:Impact.2022, Holzer.Franke.ea:Current.2022, 
Kratschmer.Sun.ea:Fully.2023, Bruder.Franzke.ea:Paramagnetic.2022,
Franzke:Reducing.2023, Bruder.Franzke.ea:Zero-Field.2023,
Franzke.Bruder.ea:Paramagnetic.2024}
were presented.
In contrast, the correlation contribution has received less
attention. That is, common approximations such as the
VWN, \cite{Vosko.Wilk.ea:Accurate.1980}
PBE, \cite{Perdew.Burke.ea:Generalized.1996}
PW92, \cite{Perdew.Wang:Accurate.1992}
B88, \cite{Becke:Correlation.1988}
or B95 \cite{Becke:Density-functional.1996} correlation
are modified and the LHF parameters are optimized by, e.g.,
thermochemical calculations on large sets of molecules.

For a straightforward applicability to arbitrary fermions,
a tailored correlation is, however, of great importance.
Notably, DFT is not restricted to electrons but can be extended 
to a many-fermion version, termed multicomponent DFT (MC-DFT).
\cite{Kreibich.Gross:Multicomponent.2001, InB-Leeuwen.Gross:2006,
Chakraborty.Pak.ea:Development.2008, Kreibich.Leeuwen.ea:Multicomponent.2008}
Most commonly, MC-DFT is used with protons, as the
respective MC-DFT approach, termed nuclear electronic orbital (NEO), 
goes beyond the established Born--Oppenheimer approximation. 
\cite{Messud:Generalization.2011, Brorsen.Yang.ea:Multicomponent.2017, 
Yu.Hammes-Schiffer:Nuclear-Electronic.2020, Pavosevic.Culpitt.ea:Multicomponent.2020, 
HammesSchiffer:Nuclearelectronic.2021}
Just like the common electronic DFT framework, MC-DFT also relies
on accurate density functional approximations. In the last two decades, electron-proton
\cite{Udagawa.Tsuneda.ea:Electron-nucleus.2014, Pak.Chakraborty.ea:Density.2007,
Chakraborty.Pak.ea:Development.2008, Chakraborty.Pak.ea:Properties.2009,
Sirjoosingh.Pak.ea:Derivation.2011,Sirjoosingh.Pak.ea:Multicomponent.2012,
Yang.Brorsen.ea:Development.2017,Brorsen.Schneider.ea:Alternative-forms.2018,
Tao.Yang.ea:Multicomponent.2019}
and electron-muon correlation functionals 
\cite{Goli.Shahbazian:Two-component.2022, 
Deng.Yuan.ea:Two-component.2023} were developed and successfully applied.
Ideally, a general density functional approximations applicable to
electrons, protons, and other fermions with similar accuracy should be
constructed. Here, the amount of data is very limited for other
fermions compared to electronic structure theory. 
Therefore, the correlation should ideally be derived in a non-empirical
way to ensure transferability.

In this work, we first show how to develop all functional parts,
i.e.\ the exchange, local mixing function, and correlation contributions
of a local hybrid functional from first principles.
Thus, the density functional approximation is designed in
an \textit{ab initio} fashion by satisfying theoretical constraints
instead of considering molecular benchmark data. Second, its performance
is assessed for various physical and chemical properties, ranging from
ground-state energies to second-order magnetic properties. Finally, a simple
extension of the new correlation functional to a multicomponent framework
is given.

\section{Theory}
\label{sec:theory}

\subsection{Local Hybrid Functionals}
\label{subsec:lhf}
Local hybrid functionals feature a fully position-dependent admixture
of exact exchange. The exchange part of the functional within an
unrestricted Kohn--Sham (UKS) framework reads
\begin{equation}
E_{\textrm{X}}^{\text{LHF}} = \int \sum_{\sigma = \alpha, \beta}
\left[\left\{1-a_{\sigma}(\vec{r})\right\} e_{\text{X}, \sigma}^{\text{DFT}}(\vec{r}) +
a_{\sigma}(\vec{r}) e_{\text{X}, \sigma}^{\text{HF}}(\vec{r})\right] ~ \textrm{d}\vec{r} 
\label{eq:lhf}
\end{equation}
where $a$ denotes the LMF, $e_{\text{X}, \sigma}^{\text{DFT}}$ the semilocal DFT
exchange energy density, and $e_{\text{X}, \sigma}^{\text{HF}}$ the exact-exchange
or Hartree--Fock (HF) exchange energy density. The latter is defined according to
\begin{eqnarray}
e_{\text{X}, \sigma}^{\text{HF}} (\Vec{r}) & = & - \frac{1}{2} \sum_{\mu \nu \kappa \lambda} P_{\mu \nu}^{\sigma} P_{\kappa \lambda}^{\sigma} ~ \chi_{\mu}^*(\vec{r}) ~ \chi_{\lambda}(\vec{r}) ~ A_{\kappa \nu}(\Vec{r}) \\
A_{\kappa \nu}(\Vec{r}) & = & \int \frac{\chi_{\kappa}^* (\Vec{r}\,') ~ \chi_{\nu} (\Vec{r}\,')}{| \vec{r} - \vec{r}\,'|} ~ \textrm{d}\vec{r}\,'
\end{eqnarray}
with the atomic orbital (AO) basis functions $\chi_{\mu}$ and the respective
AO spin density matrices $P_{\mu \nu}^{\sigma}$. Real-valued AO basis functions
are commonly employed and we drop the complex conjugation in the following.
In practice, Eq.~\ref{eq:lhf} is most easily evaluated with a seminumerical
integration scheme \cite{Plessow.Weigend:Seminumerical.2012} and a
common LMF, i.e.\ a spin-independent LMF, is chosen to include spin polarization.
\cite{Arbuznikov.Kaupp:Importance.2012}
This way, the resulting exchange potential follows as
\cite{Maier.Arbuznikov.ea:Local.2019}
\begin{equation}
\begin{split}
V_{\text{X}, \mu \nu}^{\text{LHF}, \sigma} = & 
-\frac{1}{2} \int a(\Vec{r}) ~ P_{\kappa \lambda}^{\sigma}
\left[ \chi_{\mu} \chi_{\kappa} A_{\nu \lambda}(\vec{r})
+ A_{\nu \kappa}(\vec{r}) \chi_{\mu} \chi_{\lambda}  \right] ~ \textrm{d}\Vec{r} \\
& \phantom{\text{$\frac{1}{2}$}} + 
\int \left\{1-a(\vec{r})\right\} \hat{d}_{\mu \nu}^{\sigma}
e_{\text{X}, \sigma}^{\text{sl}} ~ \textrm{d}\Vec{r} \\
& \phantom{\text{$\frac{1}{2}$}} + \int \hat{d}_{\mu \nu}^{\sigma} ~ a(\vec{r}) \left[
e_{\text{X}, \sigma}^{\text{HF}} - e_{\text{X}, \sigma}^{\text{sl}} \right] ~ \textrm{d}\Vec{r}
\label{eq:lhf-x}
\end{split}
\end{equation}
with the potential operator
\begin{equation}
\hat{d}_{\mu \nu}^{\sigma} = \sum_{Q \in \mathcal{Q}}
\int \frac{\partial Q (\Vec{r}\,')}{\partial P_{\mu \nu}^{\sigma}}
\frac{\partial}{\partial Q (\Vec{r}\,')} ~ \textrm{d}\Vec{r}\,'
\end{equation}
and $\mathcal{Q} = \{\rho_{\sigma}, \vec{\nabla} \rho_{\sigma},
\tau_{\sigma}, \vec{j}_{\text{p}, {\sigma}}, \dots \}$.
That is, $\mathcal{Q}$ collects all required variables, including the spin density 
$\rho_{\sigma}$, its gradient $\vec{\nabla} \rho_{\sigma}$, the kinetic energy
density $\tau_{\sigma}$, and the paramagnetic current density $\vec{j}_{\text{p}, {\sigma}}$. 
The latter two variables are defined according to
\begin{eqnarray}
\tau_{\sigma} & = & \phantom{-} \frac{1}{2m} \sum_j | \Vec{\nabla} \varphi_{j, \sigma}|^2 
\label{eq:tau} \\
\vec{j}_{\text{p}, \sigma} & = & - \frac{\text{i}}{2 m} \sum_j
\left( \varphi_{j, \sigma}^* \vec{\nabla} \varphi_{j, \sigma}  - \varphi_{j, \sigma} \vec{\nabla} \varphi_{j, \sigma}^* \right)
\end{eqnarray}
with $\varphi_{j, \sigma}$ denoting Kohn--Sham spin orbitals, and $m$
denoting the mass of the fermion, with $m = 1$\,a.u.\ for an electron.
Herein, the paramagnetic current density is only needed for current-carrying
states, \cite{Dobson:Alternative.1993, Becke:Current.2002, Tao:Explicit.2005}
i.e.\ for the description of excited states, \cite{Bates.Furche:Harnessing.2012, 
Holzer.Franzke.ea:Assessing.2021, Grotjahn.Furche.ea:Importance.2022,
Grotjahn.Furche:Gauge-Invariant.2023}
magnetic properties, \cite{Reimann.Ekstrom.ea:importance.2015,
Schattenberg.Kaupp:Effect.2021, Holzer.Franzke.ea:Assessing.2021,
Franzke.Holzer:Impact.2022, Pausch.Holzer:Linear.2022}
or spin--orbit coupling. \cite{Holzer.Franke.ea:Current.2022, Franzke.Holzer:Current.2024, 
Bruder.Franzke.ea:Zero-Field.2023, Bruder.Franzke.ea:Paramagnetic.2022}
Integration over $\vec{r}\,'$ is carried out analytically, while the integration
with respect to $\vec{r}$ is performed on a finite grid. \cite{Holzer:improved.2020}
We note in passing that the exchange part of a local hybrid may further include
a so-called calibration function to consider the ambiguity of the exchange energy
densities. \cite{Tao.Staroverov.ea:Exact-exchange.2008, Arbuznikov.Kaupp:Towards.2014,
Cruz.Lam.ea:Exchange-Correlation.1998, Burke.Cruz.ea:Unambiguous.1998}

\subsection{Local Exchange Enhancement Factor}
\label{susec:exchange}

Exchange functionals are defined in terms of a suitable enhancement factor $F_{\text{X}}$ 
to construct the local exchange from the exchange energy per particle of the uniform gas.
Hence, the exchange energy reads
\begin{equation}
E_{\text{X}}^{\text{DFT}} = \int \sum_{\sigma = \alpha, \beta} F_{\text{X}} \left(\rho_{\sigma}, \Vec{\nabla} \rho_{\sigma}, \tau_{\sigma}; \vec{r} \right) 
\cdot \epsilon_{\text{X}}^{\text{unif}} (\rho_{\sigma}; \vec{r}) ~ \text{d} \vec{r}
\end{equation}
with the exchange energy per electron of the uniform gas given by
\begin{equation}
\epsilon_{\text{X}}^{\text{unif}} (\rho_{\sigma}; \vec{r}) = -\frac{3}{4\pi} (3 \pi^2 \rho_{\sigma})^{1/3}
\end{equation}
In the present work, the enhancement factor is a general functional of the
density $\rho_{\sigma}$, the gradients $\Vec{\nabla} \rho_{\sigma}$, 
and the kinetic energy density $\tau_{\sigma}$. Higher-order derivatives which are 
typically used for the calibration function with local hybrids 
\cite{Maier.Haasler.ea:New.2016} are not considered.
For clarity, we use $n = \rho_{\alpha} + \rho_{\beta}$ for the particle
or total density and $\rho_{\sigma}$ for the spin densities.
To define the enhancement factor, we will further use common definitions of
density-dependent variables. 
The dimensionless density gradient $s$ is defined as
\begin{equation}
s_{\sigma} = |\vec{\nabla} \rho_{\sigma}| / \left( 2(3\pi^2)^{1/3} \rho_{\sigma}^{4/3} \right)
\end{equation}
$\tilde{q}$ is defined as
\begin{equation}
\tilde{q}_{\sigma} = \frac{9}{20} \left( \alpha_{\sigma} - 1 \right) + \frac{2}{3} p_{\sigma}
\end{equation}
with the dimensionless variables
\begin{equation}
    \alpha_{\sigma}  =  \left(\tau_{\sigma} - \tau_{\sigma}^{\text{vW}} \right)/ \tau_{\sigma}^{\text{unif}}
\end{equation}
Further, the well-known variables
\begin{equation}
\tau_{\sigma}^{\text{vW}} = {|\Vec{\nabla} \rho_{\sigma}|^2}/{ \left(8 m \rho_{\sigma} \right)}
\end{equation}
and
\begin{equation}
\tau_{\sigma}^{\text{unif}} = 3/(10m) (3\pi^2)^{2/3} \rho_{\sigma}^{5/3}
\end{equation}
denote the von-Weiz\"acker kinetic energy density and the Thomas--Fermi kinetic
energy density of the uniform electron gas, respectively.

In the TMHF functional,\cite{Holzer.Franzke:Local.2022} the 
exchange functional is derived by re-parametrizing the Tao--Mo (TM) meta-generalized
gradient approximation exchange.\cite{Tao.Mo:Accurate.2016} 
We revise this, and adapt the slowly-varying part of the strongly-constrained 
appropriately normed (SCAN) exchange functional
\cite{Sun.Ruzsinszky.ea:Strongly.2015}
\begin{equation}
F_{\text{X}, \sigma}^{\text{SC}}=  1 + \kappa - {\kappa} / \left(1+\frac{x}{\kappa} \right) 
\end{equation}
with
\begin{equation}
\begin{split}
    x = & \mu_{\text{GE}} p \left( 1 + \frac{|b_4| p}{\mu_{\text{GE}}} \exp \left[ \frac{-|b_4| p}{\mu_{\text{GE}}}\right] \right) \\
        & + \left( b_1 p + b_2 (1-\alpha) \exp \left[ -b_3(1-\alpha)^2 \right] \right)^2
\end{split}
\end{equation}
where $\mu_{\text{GE}} = 10/81$ and $p = s^2$, and the parameters $b_1$ to $b_4$ are defined as 
in Ref.~\citenum{Sun.Ruzsinszky.ea:Strongly.2015}. The parameter $\kappa$ 
in the resummation is set to $\kappa = 0.174$, the assumed optimally tight bound
for any value of $\alpha$.\cite{Perdew.Ruzsinszky.ea:Gedanken.2014}
Similar choices based on the Lieb--Oxford bounds have been made for PBE and TPSS
functionals earlier.\cite{Perdew.Burke.ea:Generalized.1996, Tao.Perdew.ea:Climbing.2003}

For the iso-orbital region, Tao and Mo have derived a suitable expression 
from the density matrix expansion (DME), yielding the enhancement factor 
\cite{Tao.Mo:Accurate.2016}
\begin{equation}
\label{eq:Fx_DME}
F_{\text{X}, \sigma}^{\text{DME}} = \frac{1}{f_{\sigma}^2} + \frac{7R_{\sigma}}{9f_{\sigma}^4}
\end{equation}
with the dimensionless functions
\begin{equation}
\label{eq:R}
\begin{split}
R_{\sigma} = & 1 + \frac{594}{54} y_{\sigma}  - {\frac{1}{\tau_{\sigma}^{\text{unif}}}} ~ \times \\
 & \left[ \tau_{\sigma} -  3\left({\lambda^2 - \lambda + 0.5} \right) 
 \left(\tau_{\sigma} - \tau_{\sigma}^{\text{unif}} - \frac{\tau_{\sigma}^{\text{vW}}}{9} \right)
 \right]
\end{split}
\end{equation}
and
\begin{equation}
    f_{\sigma} =  1 + 10\frac{70}{27}p_{\sigma} + \beta_X p_{\sigma}^2
\end{equation}
$y_{\sigma}$ has been defined as
\begin{equation}
y_{\sigma} = (2\lambda - 1)^2 p_{\sigma}
\end{equation}
We note in passing that these equations are derived with the help of the
second-order gradient expansion of the kinetic energy density.
\cite{Brack.Jennings.ea:On.1976, Perdew.Kurth.ea:Accurate.1999}
The parameters $\lambda = 0.6866$ and $\beta_{1e} = 79.873$ were then fitted to the
hydrogen atom by minimizing $\lambda$ under the constrained of $F_{\text{X}}^{\text{SC}}$
being a strictly monotonic increasing function of $s^2$.\cite{Tao.Mo:Accurate.2016} 
Contrary, in our new approach any one electron density will be exact for 
the local exchange part. Therefore, another way of determining the optimal 
parameter $\beta_{\text{X}}$ for general applications needs to be used, 
which will be outlined later.
Contrary, the value for $\beta_{1e} = 79.873$ can be kept for
other purposes, as for example in the construction
of pure local density based quantities as the LMF or correlation parts.

From the enhancement factors in the slowly varying and iso-orbital limit,
the final exchange enhancement $F_{\text{X}}$ is obtained using the same
interpolation function as in the SCAN functional,
\begin{equation}
\label{eq:final_exchange}
F_{\text{X}, \sigma} = f_{\text{X}}(\alpha) F_{\text{X}}^{\text{iso}} + \left(1 - f_{\text{X}}(\alpha) \right)F_{\text{X}, \sigma}^{\text{SC}}
\end{equation}
with the interpolation function 
\begin{equation}
\label{eq:xinter}
f_{\text{X}} (\alpha)= 
\begin{cases}
f_{\text{X}, \text{Chebyshev}}(\alpha) & 0 \le \alpha < 1 \\
0 & \text{else}
\end{cases}
\end{equation}
being different from the ones previously used by the TM and TMHF functionals.
The Chebyshev polynomials are fit to the function 
$\exp\left[ \left(-c_{\text{X}} \alpha \right) / \left(1-\alpha \right) \right]$
in the interval $[0,1]$, requiring that $f_{\text{X}}(\alpha=0) = 1$
and $f_{\text{X}}(\alpha=1) = 0$ after the optimal value of $c_{\text{X}}$
had been determined. 

\subsection{Local Mixing Function}
\label{subsec:lmf}

We start with the construction of a local mixing function 
from the correlation length\cite{Johnson:Local-hybrid.2014, Holzer.Franzke:Local.2022}
\begin{align}
\label{eq:zdme}
z^{\text{DME}}_{\sigma \sigma'} = & \left( |{U}_{\sigma}^{\text{DME}}|^{-1} + |U_{\sigma'}^{\text{DME}}|^{-1} \right)
\end{align}
as suitable indicator.
The hole functions $U$ are obtained as
\begin{equation}
\label{eq:udme}
{U}^{\text{DME}}_{\sigma} = c_{\text{F}} \left[(1+\zeta)\rho_{\sigma} \right]^{1/3}
\left( \frac{1}{{f}^2} + \frac{7{R}}{9{f}^4} \right)
\end{equation}
with $c_{\text{F}}=3/8 \cdot 4^{2/3} (3/\pi)^{1/3}$ and the relative spin polarisation
\begin{equation}
    \zeta=(\rho_{\sigma} - \rho_{\sigma'})/n
\end{equation}
$R$ and $f$ have already been defined in Eqs.~\ref{eq:R}, 
and the one-electron (high-density) values for $\lambda$ and $\beta_{1e}$
are used.\cite{Tao.Mo:Accurate.2016, Holzer.Franzke:Local.2022}
While the correlation length yields a reasonable 
asymptotic behavior, it exhibits deficiencies in slowly-varying 
and core regions. To remedy those, we therefore
enhance the correlation length, leading to a LMF of the form
\begin{equation}
a = 1 - \exp \left[ - c_{\text{1L}} \left(\Phi^{\text{SC}} + \Phi^{\text{iso}}\right) z_{\alpha \beta}^{\text{DME}}\right]
\end{equation}
where $\Phi$ mark the z-LMF enhancement functions and $c_{\text{1L}}$ is a parameter
to be optimized later.

In the slowly varying region, we exploit that the second-order gradient 
expansion of the correlation yields suitable information about the rate
at which the correlation vanishes.
We assume that a suitable switching to exact exchange should therefore
take place at the same rate, yielding\cite{Wang.Perdew:Spin.1991, 
Perdew.Burke.ea:Generalized.1996}
\begin{equation}
    \Phi^{\text{SC}} = \left[1-\left(\frac{\tau^{\text{vW}}}{\tau} \right)^2 \right] \left( c_{\text{2L}} + c_{\text{3L}} H \right)
\end{equation}
with $H = \beta(r_{\text{s}}) \phi^3 t^2$ being the second-order gradient
expansion of the correlation energy of a uniform electron gas
in the slowly-varying limit.
\cite{Perdew.Burke.ea:Generalized.1996} 
Further, $r_{\text{s}}$ denotes the local
Seitz radius from $n = 3/(4 \pi r_{\text{s}}^3) = k_{\text{F}}^3/ (3 \pi^2$). $\phi$
is a spin scaling factor \cite{Wang.Perdew:Spin.1991}
and $t$ is a dimensionless density gradient.
\cite{Perdew.Burke.Wang:Generalized.1996}
We note in passing that the kinetic energy densities $\tau^{\text{vW}}$ 
and $\tau$ are now obtained from the total density, and not 
from the spin density as in the exchange enhancement factor.
To emphasize this, the spin index $\sigma$ has been dropped
in the corresponding quantities.
The spin scaling factor $\phi$ is defined according to
\begin{equation}
    \phi = \frac{1}{2} \left[ \left( 1 + \zeta \right)^{2/3} + \left( 1 - \zeta \right)^{2/3} \right]
\end{equation}
with the relative spin polarization $\zeta$. The dimensionless density
gradient $t$ reads
\begin{equation}
    t   = | \vec{\nabla} n | / \left( 2 k_{\text{s}} \phi n \right) 
\end{equation}
based on the local Thomas--Fermi screening wave number
\begin{equation}
    k_{\text{s}} = \left( 4 k_{\text{F}} / \pi \right)^{1/2}
\end{equation}
We note in passing that the density gradient $s$ refers to
the scale of the local Fermi wavelength, $ 2 \pi / k_{\text{F}}$,
whereas $t$ corresponds to the scale of the local Thomas--Fermi
screening length, $1/k_{\text{s}}$.
To construct $H$, we use the revTPSS definition of
$\beta(r_{\text{s}}) = 0.066725 (1+0.1 r_{\text{s}})/(1+0.1778 r_{\text{s}})$ 
outlined in Ref.~\citenum{Perdew.Ruzsinszky.ea:Workhorse.2009}. 

In the iso-orbital limit, which encompasses the core region, 
a suitable LMF enhancement must at least cancel the negative cusp of 
$z^{\text{DME}}$ at the position of the nucleus to yield the overall
correct scaling to the high-density limit.
A suitable enhancement of the correlation length $z$ that fulfils this constraints 
and scales correctly under uniform coordinate scaling is
given by
\begin{equation}
    \Phi^{\text{iso}} = \left(\frac{\tau^{\text{vW}}}{\tau} \right)^2 \left( 1 + c_{\text{3L}}  H t^2 n^{2/3} \right)
\end{equation}

Using the MacLaurin series of an exponential function, it is
straightforward to show that for the slowly varying region, 
where $z \sim 1$ when approaching the high density limit, 
the correct $\gamma^{-1}$ scaling is obtained for the 
leading term of the complement of the LMF under uniform coordinate scaling.
In the iso-orbital limit, where the erroneous scaling of $z~\gamma$
is observed, again the leading term exactly cancels this, yielding an
overall $\gamma^{-1}$ scaling of the LMF as the high density limit 
is approached under uniform coordinate scaling.

We note that $a$ is a so-called common LMF, as it incorporates both spin
contributions. That is, the LMF is equal for both spins.

\subsection{Correlation in Local Hybrid Functionals}
\label{susec:correlation}

The PBE functional has been shown to be generally suitable under most
circumstances in the slowly varying limit. To deal with the increased amount 
of exact exchange in our local hybrid ansatz, we therefore suggest to 
use the localized version of the PBE functional \cite{Constantin.Fabiano.ea:Semilocal.2012}
\begin{equation}
\label{eq:corr_pbe}
 E_{\text{C}}^{\text{SC}} = E_{\text{C}}^{\text{PBEloc}}
\end{equation}
or of the B95 correlation energy
\begin{equation}
\label{eq:corr_b95}
 E_{\text{C}}^{\text{SC}} = E_{\text{C}}^{\text{B95}}
\end{equation}

However, any form of PBE is unsuitable for the limiting iso-orbital case, 
being unable to yield (nearly) vanishing one electron energies, or 
obtain the correct correlation energy in the low-density strongly 
interacting limit. B95 contrary does not yield the correct
values in the high-density iso-orbital limit.
A more suitable form in this limit must therefore be constructed 
from knowledge of the behavior of electrons in the iso-orbital limit. 
\begin{enumerate}
\item For a non-degenerate reference, 
the electron correlation for two-electron systems is approaching a 
finite limited value. Additionally, correlation energies in physical systems 
are only weak functions of the density, i.e.\ correlation 
in H$^{-}$, He, and in the limit of $Z\rightarrow \infty$ are only weakly 
density dependent.\cite{Umrigar.Gonze:Accurate.1994}
\item In the low density limit of the iso-orbital region, 
the correlation energy becomes independent of spin polarization.
\item Inter-fermion correlation energies in multicomponent DFT
have recently also been shown to only be weakly dependent on the density.
\cite{Holzer.Franzke:Beyond.2024}
\end{enumerate}
In addition to these observations, the correlation function
should only use occupied KS orbitals throughout for simplicity.
\cite{Becke:Perspective.2014}

Following the approach of Becke, we use a coupling
strength integration to derive a valid correlation functional.
\cite{Becke:Correlation.1988}
We propose a coupling strength integrand
\begin{equation}
\label{eq:integrand}
 h^{\alpha \beta}_{\text{C}, \lambda} (\vec{r},u) = 
 \frac{2}{\pi} \text{arctan}\left(c_{\text{2C}} \tilde{z}_{\alpha \beta} \lambda \right) \rho_{\beta} 
 \left(u-\tilde{z}_{\alpha \beta} \right) F \left(\gamma_{\alpha \beta} u \right)
\end{equation}
with $\gamma_{\alpha \beta}$ being defined as 
\begin{align}
    \gamma_{\alpha \beta} = & \frac{I_3}{\tilde{z}_{\alpha \beta} I_2} \\
    I_n = & \int_0^{\infty} x^n F(x) \textrm{d}x
\end{align}
according to Eqs.~(37) and (38) of Ref.~\citenum{Becke:Correlation.1988}.
$\tilde{z}$ is a scaled version of the correlation length $z^{\text{DME}}$,
i.e.
\begin{equation}
\tilde{z}_{\alpha \beta} = c_{\text{1C}} ~ z_{\alpha \beta}^{\text{DME}}
\end{equation}
The parameters $c_{\text{1C}}$ and $c_{\text{2C}}$ will be subject
to a later optimization.
The damping function $F(x)$ is equivalent to the choices presented 
in Eq.~(48) of Ref.~\citenum{Becke:Correlation.1988} and provides
a cutoff for the correlation hole when $u$ becomes large.
Note that the contribution of the exchange hole vanishes for the opposite-spin
case. Hence, $ h^{\alpha \beta}_{\text{C}, \lambda} (\vec{r},u)
= h^{\alpha \beta}_{\lambda} (\vec{r},u)$.
Proceeding as outlined by Becke, the potential energy of correlation at
a given coupling strength, $U^{\alpha \beta}_{\lambda}$, is obtained as
\begin{equation}
\begin{split}
 U^{\alpha \beta}_{\lambda} = & \frac{1}{\pi} \int \int 
    \frac{\rho_{\alpha}(\vec{r})}{u} 
    h^{\alpha \beta}_{\lambda} (\vec{r},u) ~ \textrm{d}u ~ \textrm{d}\vec{r} \\
    = & 4 \frac{I_2^2}{I_3^3} \left( I_2^2 - I_1 I_3 \right)
    \int \rho_{\alpha} \rho_{\beta} \tilde{z}^3_{\alpha \beta} 
    \text{arctan}\left(c_{\text{2C}} \tilde{z}_{\alpha \beta} \lambda \right) ~ \textrm{d}\vec{r}
\end{split}
\end{equation}
Subsequently, integration over the coupling strength $\lambda$ is carried out, 
leading to the final form of the correlation functional in the iso-orbital
limit according to
\begin{equation}
\label{eq:chyf_corr}
\begin{split}
  & E_{\text{C}}^{\alpha \beta, \text{iso}} = \int  \int_0^1 U^{\alpha \beta}_{\lambda} ~ \textrm{d}\lambda \textrm{d} \vec{r} = \int_{-\infty}^{\infty} 
  4 \frac{I_2^2}{I_3^3} \left( I_2^2 - I_1 I_3 \right) \times \\
   & \frac{\rho_{\alpha} \rho_{\beta}}{2 c_{\text{2C}}} \left[\tilde{z}_{\alpha \beta} \text{ln}\left(1 + c_{\text{2C}}^2 \tilde{z}_{\alpha \beta}^2\right) 
   - 2 c_{\text{2C}} \tilde{z}_{\alpha \beta}^2 \text{arctan}\left(c_{\text{2C}} \tilde{z}_{\alpha \beta}\right) \right] \textrm{d} \vec{r}
\end{split}
\end{equation}
The prefactor is evaluated as
\begin{equation}
    4 \frac{I_2^2}{I_3^3} \left( I_2^2 - I_1 I_3 \right) \approx 0.5
\end{equation}
in a straightforward manner following Ref.~\citenum{Becke:Correlation.1988}.
It differs from Becke's suggested values simply by the prefactor of
$2 \pi^{-1}$ introduced to normalize the arctan function.
Eq.~\ref{eq:chyf_corr} provides an interesting result, outlining
the correlation energy as a difference between two separate functions.
Unlike the original ansatz, Eq.~\ref{eq:chyf_corr} converges to a finite
limit in the high density case where $\tilde{z}_{\alpha \beta} \rightarrow 0$.

We note in passing that in the extreme low-density limit, the correlation
energy in the iso-orbital region $E_{\text{C}}^{\text{iso, low}}$
becomes the spin-averaged other-spin correlation energy.
To obtain the latter, $\tilde{z}_{\alpha \beta}$ is replaced by
\begin{equation}
\label{eq:zbar}
\underline{\tilde{z}} = 2 c_{\text{1C}} \left( |\underline{U}^{\text{DME}}|^{-1} \right)
\end{equation}
with the averaged hole approximation
\begin{equation}
\label{eq:udme_average}
\underline{U} = c_{\text{F}} \left[\frac{n}{2} \right]^{1/3} \left( \frac{1}{\underline{f}^2} + \frac{7\underline{R}}{9\underline{f}^4} \right)
\end{equation}
Note that for any spin-unpolarized system, $\underline{\tilde{z}} = \tilde{z}$.
Subsequently, in the evaluation of the functions $\underline{f}$ and $\underline{R}$, 
also the spin averaged quantities are used, i.e.\ $\rho_{\sigma} \rightarrow  n/2$.
This limit is also important in the case of the interaction of
different fermions, where only the interaction between the 
averaged fermion densities is accounted for.

To yield a correlation functional that is also valid 
for the uniform electron gas, we interpolate between the PBE or B95 
correlation functionals and the derived iso-orbital energy in an approach similar to 
Ref.~\citenum{Sun.Ruzsinszky.ea:Strongly.2015} by using the interpolation
\begin{equation}
    \epsilon_{\text{C}} (\vec{r}) = f_{\text{C}} (\alpha,\vec{r}) \epsilon_{\text{C}}^{\text{iso}} (\vec{r}) + \left[ 1 - f_{\text{C}}(\alpha,\vec{r}) \right] \epsilon_{\text{C}}^{\text{SC}}(\vec{r})
\end{equation}
where $f_{\text{C}}(\alpha)$ is given by the function
\begin{equation}
\label{eq:cinter}
    f_{\text{C}}(\alpha) = 
    \begin{cases}
        f_{\text{C}, \text{Chebyshev}}(\alpha) & 0 \le \alpha < 1 \\
        0 & \text{else}
    \end{cases}
\end{equation}
and $\epsilon_{\text{C}}$ is the correlation energy per electron.
The Chebyshev polynomials are fit to the function 
$\exp\left[ \left(-5 \alpha \right) / \left(1-\alpha \right) \right]$
in the interval $[0,1]$, requiring that $f_{\text{C}}(\alpha=0) = 1$
and $f_{\text{C}}(\alpha=1) = 0$. A large prefactor of 5 is chosen
to make $\partial E_{\text{C}} / \partial \alpha$ sizable near $\alpha \approx 0$ 
in the low-density, strongly interacting limit. This will lead to a pronounced 
current-density response in the presence of a magnetic perturbation.
The resulting correlation energy, while rather complicated 
formally, is numerically robust. It furthermore scales
correctly to the high-density iso-orbital limit, 
and recovers the correct LDA correlation expression
in the slowly varying region.

Finally, we note that the same-spin correlation energy
vanishes in the iso-orbital limit, and therefore no separate
same-spin correlation for this region is included in
our correlation functional.

\subsection{Optimization of Parameters}

Now we need to determine the seven parameters occurring in our
exchange ($c_{\text{X}}$, $\beta_{\text{X}}$ ), LMF ($c_{\text{1L}}$, $c_{\text{2L}}$, $c_{\text{3L}}$), 
and correlation models ($c_{\text{1C}}$, $c_{\text{2C}}$). While it is possible to simply
optimize all of them using thermochemical datasets, we lean towards more
general ways of doing so. Conveniently, in our correlation model
only parameters from the iso-orbital limit are needed, where it
is known that Hartree--Fock is an excellent approximation.
We therefore optimize $c_{\text{1C}} = 0.875$ and $c_{\text{2C}} = 0.38$ to fit the
correlation energies of the two-electron systems
H$^{-}$, He, Be$^{2+}$, Ne$^{8+}$, and Hg$^{78+}$, for
which accurate values are known.\cite{Umrigar.Gonze:Accurate.1994}

Next, we fit $\beta_{\text{X}} = 117.0$ and $c_{\text{1L}} = 0.18$ and 
$c_{\text{3L}} = 0.10$ to the total energies of the same systems, respecting 
that in our LHF exchange and correlation can no longer be strictly
separated. $c_{\text{1L}}$ and $c_{\text{3L}}$ describe the rate at which 
exact exchange is incorporated depending on the inhomogenity
of the system. $\beta_{\text{X}}$ contrary is linked to the
exchange enhancement factor, and larger values of 
$\beta_{\text{X}}$ lead to a faster damping of the exchange energy
in inhomogeneous regions. It is detrimental to understand that the
need to optimize $\beta_{\text{X}}$ arises from the neglect of a 
gauge transformation. If instead a gauge-correction is used, 
$\beta_{\text{X}} = \beta_{1e} = 79.873$, 
and the gauge transformation must be chosen 
as to recover the total energies of the two-electron systems.
Two major problems of a possible gauge transformation still arise.
First, an optimal gauge transformations aligns exact and semilocal exchange, 
making the determination any mixing parameters between semilocal and exact 
exchange difficult. Second, gauge transformations involve higher derivatives 
of the density, i.e.\ Laplacians and even Hessians. This often prevents 
convergence of iterative procedures, leading to numerical instabilities 
in functionals using gauge transformations. We therefore currently neglect 
gauge transformations and instead re-optimized the DME, as we know of no
adequate yet stable formulation suitable for our exchange model. 
While developing such a transformation would be helpful, this goes well 
beyond the scope of this already quite extensive work.

Note that for any two-electron system $\alpha = 0$, 
therefore $c_{\text{X}}$ and $c_{\text{2L}}$ cannot be optimized using them.
For $c_{\text{X}}$, it has, however, been shown that this parameter
is crucial in the construction of ultra-nonlocal metaGGA exchange
models, as for example the TASK exchange functional.\cite{Aschebrock.Kummel:Ultranonlocality.2019}
We therefore fit $c_{\text{X}} = 0.83$ for the pure exchange functional to the 
reported TASK polarizabilities of hydrogen chains with 4 to 18 atoms.\cite{Aschebrock.Kummel:Ultranonlocality.2019}
The final parameter $c_{\text{2L}}$ determines the damping of correlation length
in the slowly varying region, and it should ideally be equal to 0.
However, previous experience from DFT as well as thermochemical 
optimization using the W4-11 and BH76 test sets hint at a value of
$c_{\text{2L}} = 0.2 $ being more optimal.

\begin{table}[b]
\centering
\caption{Coefficients $b_v$ for the Chebyshev polynomial used to
fit the interpolation functions $f_{\text{X}}$ and $f_{\text{C}}$ 
of Eqs.~\ref{eq:xinter} and \ref{eq:cinter}.}
    \label{tab:cheby}
\begin{tabular}{c
S[table-format=-1.7]
S[table-format=-1.7]}
\toprule
\text{$b_v$} & \text{$f_\text{X}$} & \text{$f_\text{C}$} \\
\midrule
0 & 0.4534882  & 0.2326471  \\
1 & -0.5505752 & -0.3999473 \\
2 & 0.0375553  & 0.2494000  \\
3 & 0.0578463  & -0.1037600 \\
4 & 0.0151324  & 0.0206610  \\
5 & -0.0043256 & 0.0037073  \\
6 & -0.0061759 & -0.0027081 \\
7 & -0.0029455 & 0.0000000 \\
\bottomrule
\end{tabular}
\end{table}

To obtain numerically stable functionals, we fitted the
interpolation functions $f_{\text{X}}$ and $f_{\text{C}}$ 
of Eqs.~\ref{eq:xinter} and \ref{eq:cinter} with
\begin{equation}
    f(\alpha) = \sum_{v=0}^7 b_v C_v(2\alpha - 1)
\end{equation}
where $C_v(x)$ represents the $v-$th Chebyshev polynomial of 
the first kind evaluated at $x$. The obtained fitting coefficients 
are given in Table~\ref{tab:cheby}.

\begin{figure}[t]
    \centering
    \includegraphics[width=1.0\columnwidth]{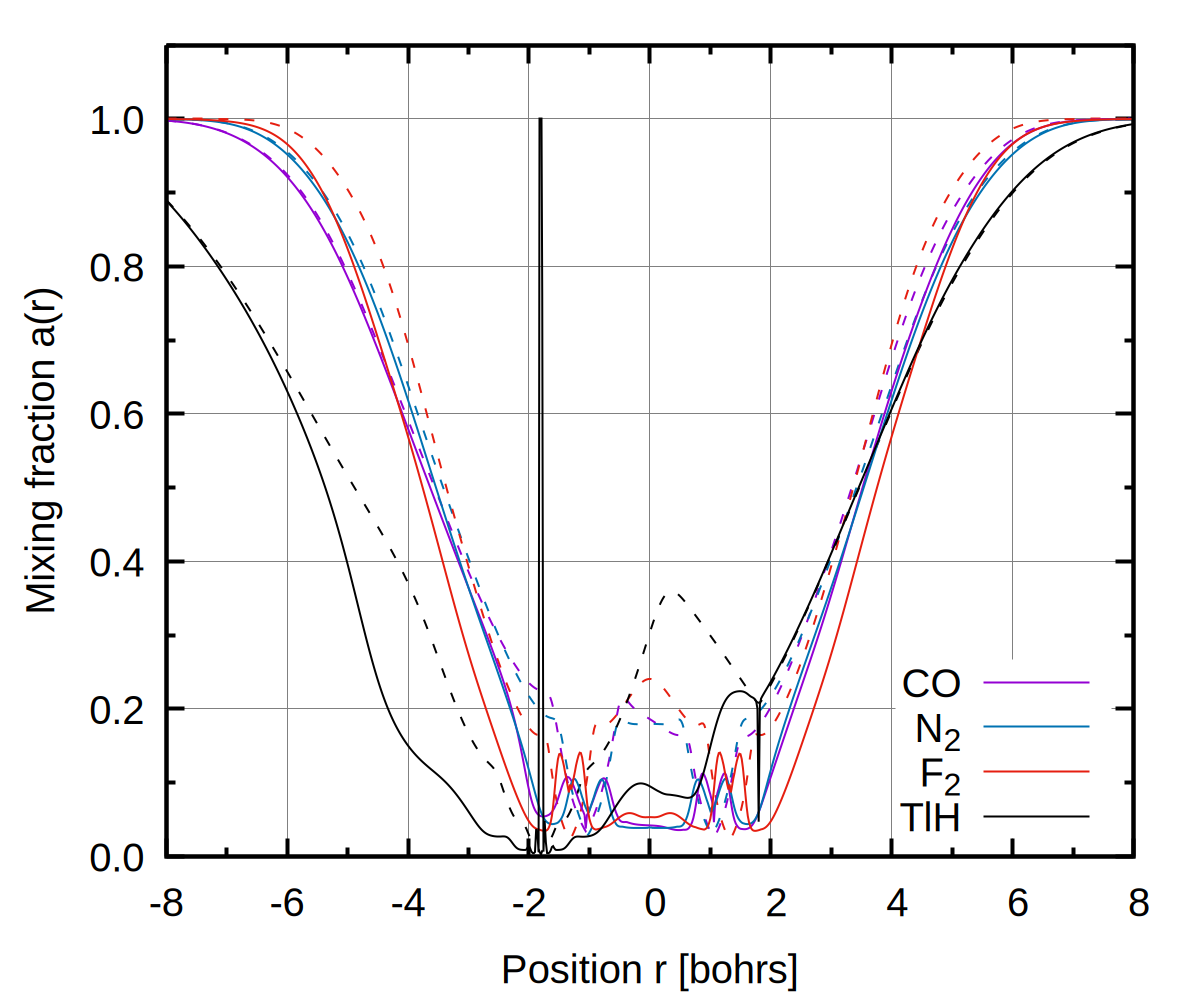}
    \caption{Local mixing function $a$ obtained from TMHF (dashed lines, Ref.~\citenum{Holzer.Franzke:Local.2022}) 
     and CHYF (solid lines, this work). 
     Properties calculated for four diatomic molecules at self-consistent
     aug-cc-pVQZ \cite{Dunning:Gaussian.1989, Kendall.Dunning.ea:Electron.1992, Woon.Dunning:Gaussian.1993}
     (H, C, N, O, F) and aug-cc-pwCVQZ-DK3 \cite{Bross.Peterson:Correlation.2014}
     (Tl) orbitals. $a$ is plotted along the internuclear axis.
     See Supporting Information for complete computational settings.}
    \label{fig:lmf}
\end{figure}

The resulting exchange-correlation functionals are denoted
CHYF and CHYF-B95 general fermions functional in this work.
As described in the previous section, the two functionals
only differ by the choice of the correlation term for the
slowly varying limit, i.e.\ locPBE vs.\ B95.
A plot of the resulting local mixing fraction $a (r)$ is shown in Fig.~\ref{fig:lmf}
and compared to the local mixing function of the TMHF functional.
The most striking change of new LMF is increases of the amount of
exact exchange at the heavy nuclei. In the bonding region, the
new LMF leads to a reduced amount of exact exchange. For instance,
for N$_2$ the TMHF LMF leads to about 18\% of HF exchange
at the center of mass, whereas the new one only leads to 
only a few percent of exact exchange being incorporated.
In the tail region, the TMHF LMF shows an more rapid increase of
$a$ but both LMFs converge to the same limit, i.e.\ $a \rightarrow 1$.

\subsection{Limitations}
The ansatz presented herein is not intended for systems with strong correlation,
i.e.\ systems with large mixing of configurations.
As shown in the B13 functional, \cite{Becke:Density.2013}
these systems require an additional term for the strong-correlation
contribution.

Further, dispersion interaction was not considered explicitly.
Thus, an extension of the presented model in this direction may be of
interest in the future. This could be done either based on the
semi-empirical D3 \cite{Grimme.Antony.ea:consistent.2010, Grimme.Ehrlich.ea:Effect.2011}
and D4 \cite{Caldeweyher.Ehlert.ea:Generally.2019} models or
based on the less empirical VV10 framework \cite{Vydrov.Van-Voorhis:Nonlocal.2010}
or even the fully parameter-free exchange-hole dipole moment (XDM) dispersion
correction. \cite{Becke.Johnson:density-functional.2005,
Becke.Johnson:Exchange-hole.2007, Otero-de-la-Roza.Johnson:Non-covalent.2013}
The first route was recently followed to obtain the D4 parameters
for TMHF and yielded encouraging results. \cite{Reimann.Kaupp:Spin-State.2023}
However, for a many-fermions framework the study of dispersion is
still in its infancy and a less empirical ansatz may be beneficial.

Finally, we note that the seminumerical implementations of LHFs (see
next subsection) are restricted to finite systems and periodic systems
are therefore beyond the scope of the present work.

\subsection{Implementation}
The new functional described herein is implemented in
TURBOMOLE \cite{Ahlrichs.Bar.ea:Electronic.1989,
Balasubramani.Chen.ea:TURBOMOLE.2020,
Franzke.Holzer.ea:TURBOMOLE.2023, TURBOMOLE}
based on the given MAPLE files, which incorporate functionalities
from Libxc. \cite{Marques.Oliveira.ea:Libxc.2012,
Lehtola.Steigemann.ea:Recent.2018, LIBXC.2023}
The functional designed in this work does not include a calibration function
and only includes the density, its gradient, and the kinetic energy density.
The latter is generalized with the current density for magnetic properties,
excited states, and spin--orbit coupling as noted above. Therefore, the
new functional is directly available for the electronic ground-state 
self-consistent field (SCF) formalism and the related expectation values, 
\cite{Bahmann.Kaupp:Efficient.2015, 
Holzer:improved.2020} analytical geometry gradients, 
\cite{Klawohn.Bahmann.ea:Implementation.2016}
excitation energies \cite{Maier.Bahmann.ea:Efficient.2015, Holzer:improved.2020,
Kehry.Franzke.ea:Quasirelativistic.2020}
and excited-state geometries \cite{Grotjahn.Furche.ea:Development.2019}
from time-dependent density functional theory (TDDFT)
as well as quasiparticle states from the Green's function $GW$
formalism, \cite{Holzer.Franzke.ea:Assessing.2021}
nuclear magnetic resonance (NMR) shifts,
\cite{Schattenberg.Reiter.ea:Efficient.2020, 
Schattenberg.Kaupp:Effect.2021, Holzer.Franzke.ea:Assessing.2021}
NMR coupling constants, \cite{Mack.Schattenberg.ea:Nuclear.2020,
Franzke.Mack.ea:NMR.2021, Holzer.Franzke.ea:Assessing.2021}
and electron paramagnetic resonance (EPR)
properties such hyperfine coupling constants,
\cite{Bruder.Franzke.ea:Paramagnetic.2022, Franzke.Yu:Hyperfine.2022}
g-tensors, \cite{Franzke.Holzer:Impact.2022, Bruder.Franzke.ea:Paramagnetic.2022,
Franzke.Yu:Quasi-Relativistic.2022}
and zero-field splitting parameters.
\cite{Bruder.Franzke.ea:Zero-Field.2023}
This way, paramagnetic NMR shifts are also directly available.
\cite{Gillhuber.Franzke.ea:Efficient.2021, Franzke.Bruder.ea:Paramagnetic.2024}
The self-consistent two-component formalism is available for
the SCF energies, \cite{Wodynski.Kaupp:Noncollinear.2020, Holzer:improved.2020,
Holzer.Franke.ea:Current.2022}
EPR properties, \cite{Wodynski.Kaupp:Noncollinear.2020, Holzer:improved.2020,
Holzer.Franke.ea:Current.2022, Franzke.Yu:Hyperfine.2022, Franzke.Yu:Quasi-Relativistic.2022}
NMR coupling constants, \cite{Franzke.Mack.ea:NMR.2021}
TDDFT excitation energies and polarizabilities,
\cite{Holzer:improved.2020, Holzer.Franke.ea:Current.2022,
Kehry.Franzke.ea:Quasirelativistic.2020}
as well as the $GW$ and Bethe--Salpeter equation (BSE) formalism.
\cite{Holzer.Klopper:Ionized.2019, Holzer:Practical.2023}
Two-component NMR shieldings and EPR g-tensors can only be calculated
with a common gauge origin. \cite{Holzer.Franke.ea:Current.2022,
Franzke.Holzer:Exact.2023}
Special relativity is either introduced with effective
core potentials \cite{Armbruster.Weigend.ea:Self-consistent.2008,
Baldes.Weigend:Efficient.2013} or all-electron approaches
such as exact two-component (X2C) theory.
\cite{Peng.Middendorf.ea:efficient.2013, Franzke.Middendorf.ea:Efficient.2018}
Here, all-electron theories such as X2C are necessary for magnetic
properties.
Furthermore, the evaluation of M\"ossbauer contact densities
with local hybrids is implemented herein, see the
Supporting Information. This means that CHYF and CHYF-B95
can be readily applied to a broad range of chemical studies.

Multicomponent DFT is available for ground-state calculations
and excitation energies. \cite{Holzer.Franzke:Beyond.2024}
For the latter, we currently neglect the inter-fermion correlation kernel.
Additionally, quasiparticle energies can be obtained based on
the Kohn--Sham solutions and the $GW$ approximation.
\cite{Holzer.Franzke:Beyond.2024}

\section{Computational Methods}
The accuracy of the new density functional approximation is assessed
for thermochemical properties such as atomization energies and
barrier heights, excitation energies, M\"ossbauer isomer shifts,
NMR spin--spin coupling constants, NMR shieldings and shifts,
magnetizabilities, as well as EPR hyperfine coupling constants.

For brevity, computational details for the benchmark studies below
are listed in the Supporting Information. Furthermore, more results for
M\"ossbauer isomer shifts or contact densities, NMR coupling constants,
NMR shielding constants, as well as EPR hyperfine coupling
constants are only presented in the Supporting Information.

\section{Results and Discussion}
\label{sec:results}

\subsection{Thermochemistry and Electronic Ground State}
\label{subsec:thermo}

\begin{figure}[t]
\centering
\includegraphics[width=1.0\linewidth]{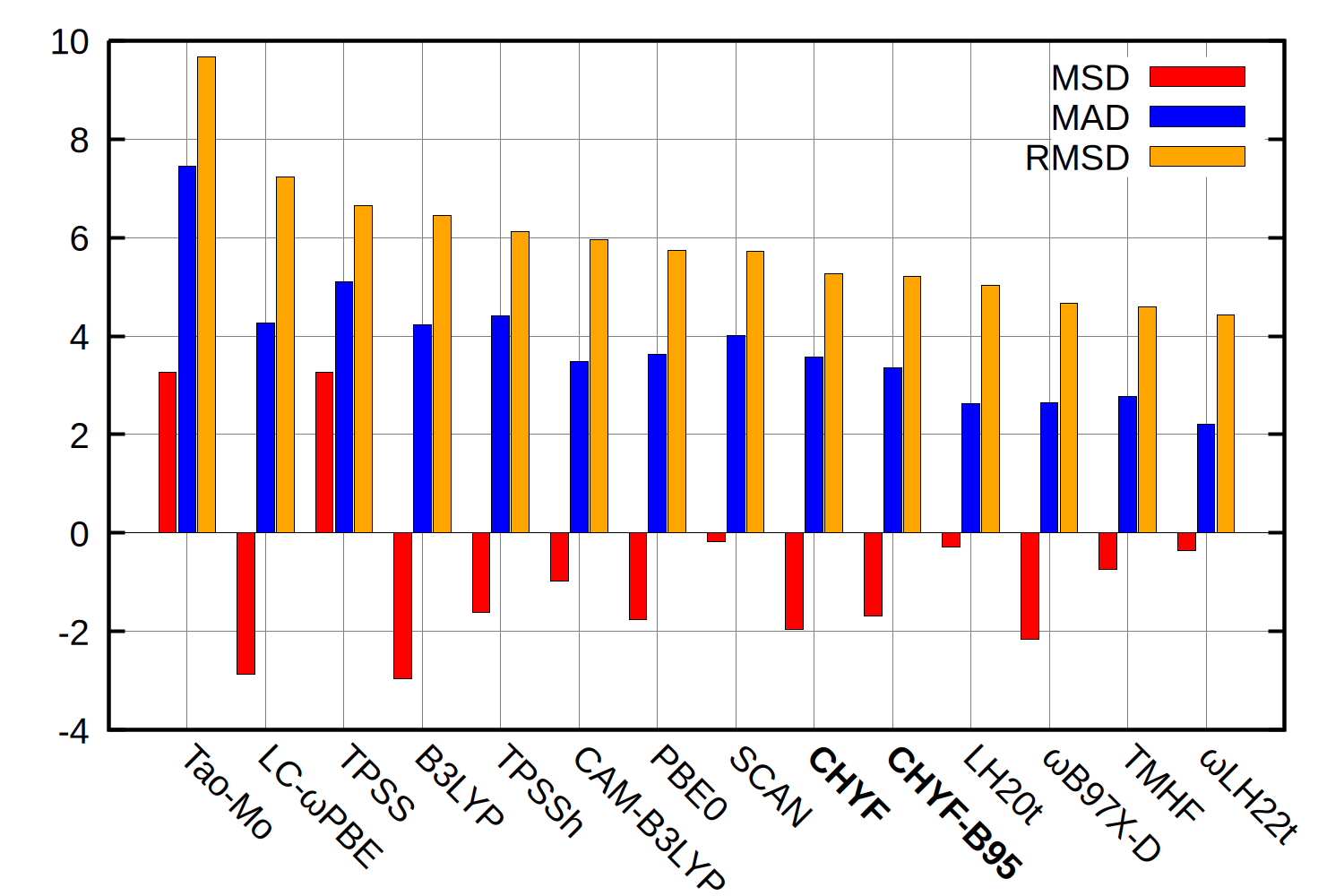}
\caption{Mean standard deviation (MSD), mean average deviation (MAD), 
and root mean square deviation (RMSD) for the atomization energies of the 
W4-11 test set. All values are in kcal/mol.}
\label{fig:w4-11}
\end{figure}

For the thermochemical W4-11 test set, \cite{Karton.Daon.ea:W4-11.2011} 
being composed of 140 atomization energies, the new local hybrid
functional is able to outperform other functionals that have been
designed with theoretically constrained satisfaction in mind.
As shown in Fig.~\ref{fig:w4-11}, CHYF generally manages to be 
better than common functionals such as
PBE0, \cite{Perdew.Burke.ea:Generalized.1996, Adamo.Barone:Toward.1999}
TPSSh, \cite{Tao.Perdew.ea:Climbing.2003, Staroverov.Scuseria.ea:Comparative.2003}
and SCAN, \cite{Sun.Ruzsinszky.ea:Strongly.2015}
with the latter yielding stellar performance for a pure
meta-GGA functional. Here, the three given functionals lead to
a slightly smaller mean signed deviation (MSD), however, the
mean average deviation (MAD) and root mean square deviation
(RMS) are larger than that of CHYF.

Even though we have carefully adapted of the correlation
energy term in Sec.~\ref{susec:correlation}, CHYF still has
a tendency of underbinding in molecular systems. 
Root mean square deviations 
are, however, comparable to thermochemically optimized local hybrids
such as LH20t, \cite{Haasler.Maier.ea:Local.2020}
and only slightly worse than those of the thermochemically
optimized range-separated hybrid {\textomega}B97X-D
\cite{Chai.Head-Gordon:Long-range.2008} and the range-separated
local hybrid {\textomega}LH22t. \cite{Fuerst.Kaupp:Accurate.2023}

For barrier heights, assessed with the BH76 test set,
\cite{Zhao.Gonzalez-Garcia.ea:Benchmark.2005,
Zhao.Lynch.ea:Multi-coefficient.2005, Goerigk.Grimme:General.2010}
CHYF performers similarly in terms of accuracy. Errors are again
smaller than those of SCAN, TPSSh, or PBE0. This therefore
confirms the findings of the atomization energies of the W4-11 test set.
The thermochemically optimized local and range-separated hybrid
functionals yield only slightly lower deviations for the BH76 test
set, as does the TMHF functional. For both test sets, the choice
of the correlation expression for the slowly varying limit,
i.e.\ PBEloc or B95, does not substantially affect thermochemically 
results.

\begin{figure}[t]
\centering
\includegraphics[width=1.0\linewidth]{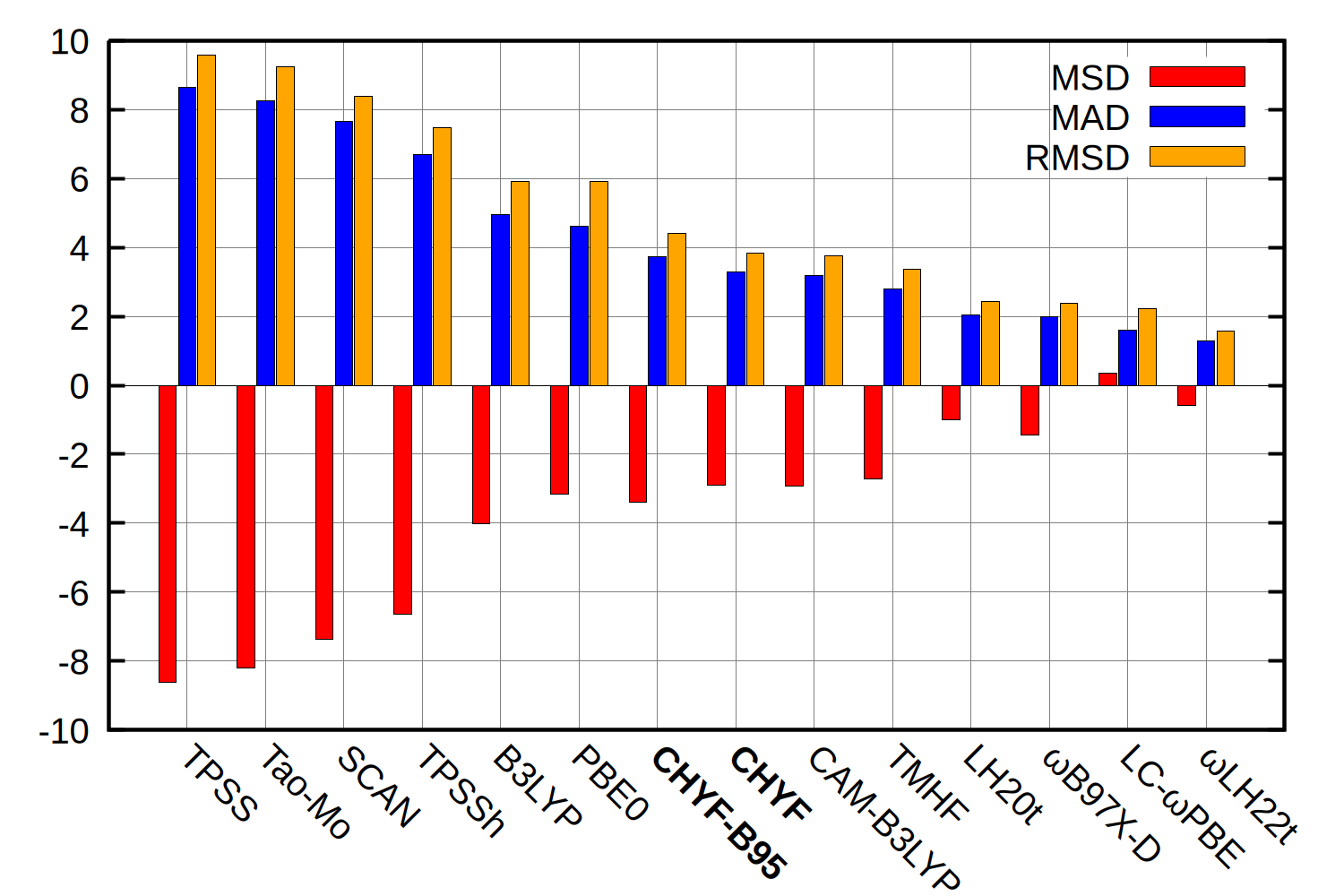}
\caption{Mean standard deviation (MSD), mean absolute deviation (MAD), 
and root mean square deviation (RMSD) for the barrier heights of the 
BH76 test set. All values are in kcal/mol.}
\label{fig:bh76}
\end{figure}

\begin{figure}[t]
\centering
\includegraphics[width=1.0\linewidth]{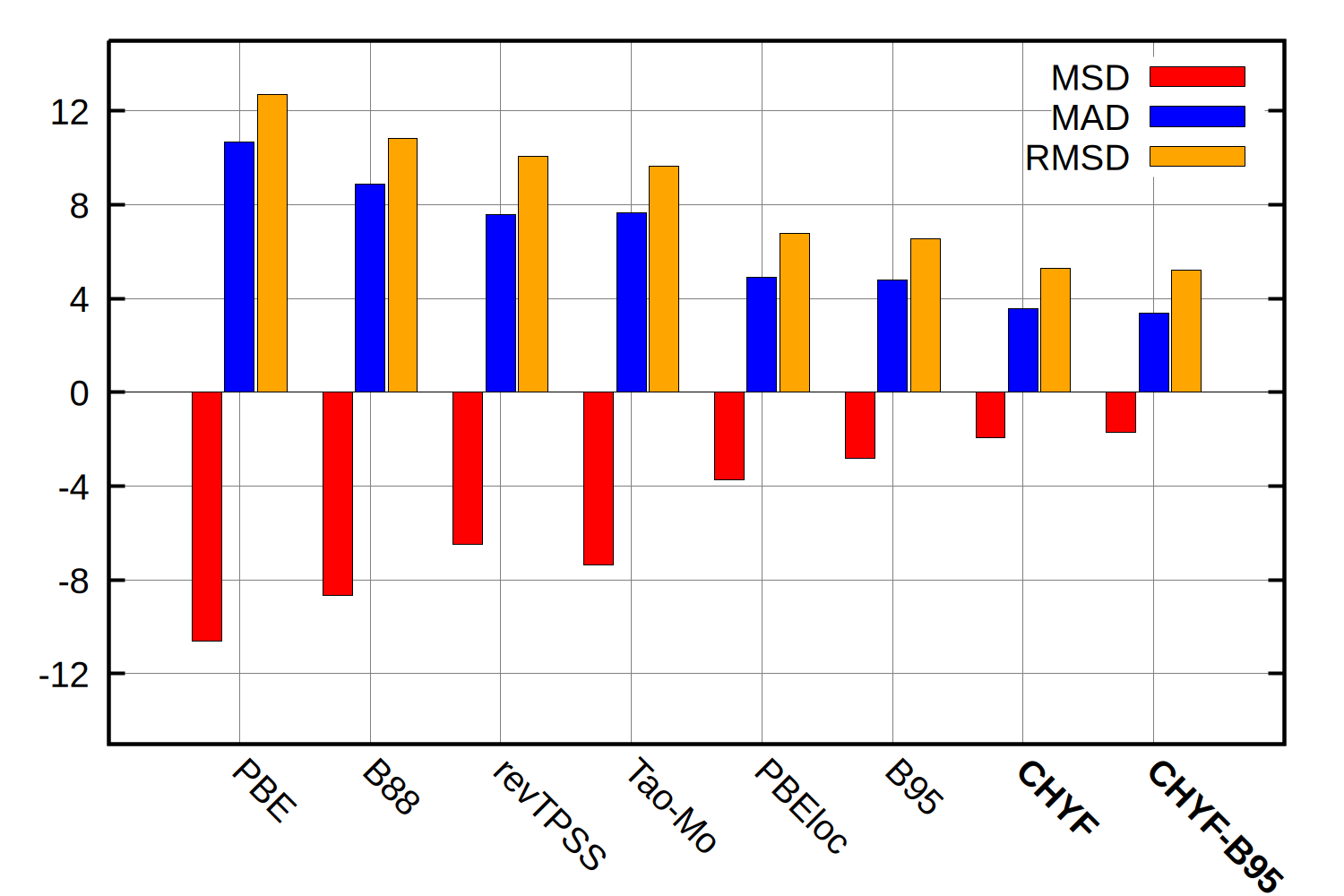}
\caption{Mean standard deviation (MSD), mean average deviation (MAD), 
and root mean square deviation (RMSD) for the atomization energies of the 
W4-11 test set for the CHYF local hybrid exchange functional
combined with different correlation functionals. All values are in kcal/mol.}
\label{fig:w4-11_corr}
\end{figure}

Overall, we deem this accuracy for thermochemistry to be clearly
sufficient for a general and transferable local hybrid, that favors a
first-principles-based construction over thermochemical optimization.
Comparing to other possible correlation functionals that lack 
the redesigned iso-orbital limit provided by Eq.~\ref{eq:chyf_corr}
further reveals the value of the latter. 
Fig.~\ref{fig:w4-11_corr} outlines the thermochemical performance
of the local hybrid exchange functional outlined in Secs.~\ref{susec:exchange}
and \ref{subsec:lmf} combined with different correlation functionals.
The less incorrect description of correlation in the iso-orbital limit
reduces the error of CHYF(-B95) by nearly 2$\,$kcal/mol when
compared to the parent B95\cite{Becke:Density-functional.1996} 
and PBEloc\cite{Constantin.Fabiano.ea:Semilocal.2012} correlation functionals. 
Other correlation functionals are less compatible with 
local hybrid exchange functionals, as outlined by the RMSD values
of revTPSS,\cite{Perdew.Ruzsinszky.ea:Workhorse.2009} Tao--Mo,\cite{Tao.Mo:Accurate.2016} 
B88,\cite{Becke:Correlation.1988} and PBE\cite{Perdew.Burke.ea:Generalized.1996} 
correlation reaching or surpassing 10$\,$kcal/mol.

\subsection{Numerical Behavior and Stability}
\label{subsec:numerical_behavior}

The numerical requirements of CHYF are further comparably low.
As outlined by Table~\ref{tab:grid_dep}, the grid dependence of the functional
is less pronounced than one may suspect from the complicated structure
arising in Sec.~\ref{sec:theory}. Already very small grids
provide sufficiently accurate results, both in terms of
relative and absolute deviations. Already with the smallest grid 1,
\cite{Treutler:Entwicklung.1995, Treutler.Ahlrichs:Efficient.1995}
results that are sufficiently converged for most purposes are 
obtained. Total energies are already accurate to $10^{-3}$\,a.u., 
which subsequently improves smoothly with increasing grid sizes.

\begin{table}[b!]
    \centering
    \caption{Grid dependence of CHYF for the W4-11 test set of 
    atomization energies. Mean absolute energy differences
    between total energies $\Delta \bar{E}$ are given with respect to
    grid 7. \cite{Treutler:Entwicklung.1995, Treutler.Ahlrichs:Efficient.1995}}
    \label{tab:grid_dep}
    \begin{tabular}{
@{\extracolsep{4pt}}
c
S[table-format=-1.3]
S[table-format=1.3]
S[table-format=1.3]
c}
    \toprule
 Grid    & \text{MSD} & \text{MAD}  & \text{RMSD} & \text{$\Delta \bar{E}$} \\
         & [kcal/mol] & [kcal/mol] & [kcal/mol] & [a.u.] \\ 
    \midrule
    7    & -1.805 & 3.346 & 5.205 & {\text{--}} \\
    6    & -1.816 & 3.348 & 5.205 & $5.44 \cdot 10^{-5}$ \\
    5    & -1.821 & 3.354 & 5.213 & $1.20 \cdot 10^{-4}$ \\
    4    & -1.716 & 3.315 & 5.175 & $2.93 \cdot 10^{-4}$ \\
    3    & -1.438 & 3.284 & 5.132 & $5.42 \cdot 10^{-4}$ \\ 
    2    & -1.690 & 3.360 & 5.211 & $8.60 \cdot 10^{-4}$ \\
    1    & -1.234 & 3.270 & 5.109 & $1.41 \cdot 10^{-3}$ \\
    \bottomrule
    \end{tabular}
\end{table}

Due to the high numerical stability, also no issues regarding 
the convergence of the SCF iterations were observed in any
calculation performed in this work.
While certainly more subjective than the results in
Table~\ref{tab:grid_dep}, 
we find that convergence is generally very smooth, with 
even challenging cases as, e.g., relativistic two-component
open-shell complexes converging very well herein.

Overall, the computational demands associated with local
hybrid functionals are substantially reduced by this
grid behavior. This allows to use small grids for
calculations without loss of accuracy. This is especially
advantageous for the exact exchange terms, which can
become a computational overhead for the seminumerical
evaluation of LHFs compared to the corresponding multigrid
approach for global or range-separated hybrids.
\cite{Neese.Wennmohs.ea:Efficient.2009, Plessow.Weigend:Seminumerical.2012,
Holzer:improved.2020, Stoychev.Auer.ea:Self-Consistent.2018, 
Holzer.Franzke.ea:Assessing.2021, Helmich-Paris.Souza.ea:Improved.2021}
The latter uses a larger grid for the semilocal DFT exchange than for
the exact exchange terms. Especially for response calculations,
very small grids are usually sufficient for the latter.
\cite{Holzer:improved.2020, Holzer.Franzke.ea:Assessing.2021,
Bruder.Franzke.ea:Zero-Field.2023, Franzke.Holzer:Exact.2023}
Therefore, we expect CHYF to be competitive to PBE0
in terms of computational costs for ``real-world''
quantum chemical studies.

\subsection{Electronic Excitation Energies with TDDFT}
\label{subsec:tddft}

The excited state test set of Ref.~\citenum{Suellen.Freitas.ea:Cross-Comparisons.2019}
is remarkable in one respect: It compiles a set of high-quality experimental
references, \textit{ab initio} data, and accounts for geometrical changes during
the excitation, as well as zero-point vibrational energy contributions.
To perform well in this test, a method must therefore be able to
describe ground and excited states reasonably well.

\begin{table}[b!]
\centering
\caption{Mean signed deviation (MSD), mean absolute deviation (MAD), 
root mean square deviation (RMSD), and maximum deviation
(Max.) for 41 excited states of 37 molecules compared to experimental
data as outlined in Ref.~\citenum{Suellen.Freitas.ea:Cross-Comparisons.2019}.
Values for other functionals are taken from Ref.~\citenum{Holzer.Franzke:Local.2022}
and Ref.~\citenum{Suellen.Freitas.ea:Cross-Comparisons.2019}.
Note that we always use the current-dependent and gauge-invariant extension
of the kinetic energy density. All values are in eV.}
\label{tab:excited_state}
\begin{tabular}{
@{\extracolsep{8pt}}
l
S[table-format=-1.3]
S[table-format=1.3]
S[table-format=1.3]
S[table-format=1.3]}
\toprule
\text{Method} & \text{MSD}    &  \text{MAD}   & \text{RMSD}  & \text{Max.} \\
\midrule
PBE       & -0.204 & 0.240 & 0.301 & 0.632   \\
TPSS      & -0.169 & 0.230 & 0.287 & 0.655   \\
B3LYP     & -0.192 & 0.265 & 0.355 & 0.828   \\
PBE0      & -0.138 & 0.273 & 0.333 & 0.734   \\
TPSSh     & -0.119 & 0.237 & 0.318 & 0.774   \\
CAM-B3LYP & -0.100 & 0.274 & 0.338 & 0.709   \\
LC-{\textomega}PBE   & -0.021 & 0.291 & 0.311 & 0.588   \\
{\textomega}B97X-D   & -0.097 & 0.273 & 0.333 & 0.693   \\
LH20t     & 0.121  & 0.258 & 0.313 & 0.658   \\
{\textomega}LH22t & 0.100 & 0.267 & 0.313 & 0.619 \\
\midrule
TMHF      & 0.002  & 0.246 & 0.276 & 0.467   \\
\textbf{CHYF}      & 0.078  & 0.246 & 0.290 & 0.566   \\
\textbf{CHYF-B95}      & 0.053  & 0.219 & 0.260 & 0.539   \\
\midrule
CC2       & 0.045  & 0.083 & 0.112 & 0.270   \\
CCSD      & 0.177  & 0.177 & 0.204 & 0.429   \\
ADC(3)    & -0.125 & 0.228 & 0.271 & 0.488   \\
CC3       & -0.011 & 0.025 & 0.036 & 0.107   \\
\bottomrule
\end{tabular}
\end{table}

As outlined by the results in Table~\ref{tab:excited_state},
this is a substantial task for density functional approximations.
Especially the root-mean-square deviation (RMSD) reveals that a 
barrier exists at 0.3\,eV. And this barrier cannot easily be overcome 
by climbing the functional ladder, as revealed by the stagnating
errors when going from (meta-)GGAs to hybrid or even local hybrid
functionals, where the RMSD is often significantly worsened when compared
to their parent functional.
Also recent local hybrid functionals obtained by extensive fitting
procedures as LH20t\cite{Haasler.Maier.ea:Local.2020} and
and the range-separated local hybrid
{\textomega}LH22t\cite{Fuerst.Kaupp:Accurate.2023} are unable 
to rectify this. Instead, they invert the general trend
of underestimating excited state energies, but no further
changes of the magnitude of errors is observed.
By using a construction based on first principles, as outlined
in this work and previously for TMHF---albeit only for exchange
in the latter case---this barrier can be overcome.
Both TMHF and CHYF cut the RMSD by approximately 20\,\% 
compared to other density functional approximations, while
also significantly cutting down on the maximum error observed.
The overall balanced description of excited states
is further emphasized by the mean signed deviation
approaching zero for both methods.
For these functionals, the accuracy levels provided by density functional
theory are within the grasp of high-level methods such as CCSD
for the first time. An odd pick of wavefunction-based
methods, as for example ADC(3), could even leave one
with no advantage over TMHF or CHYF.
We, however, admit that ADC(3) is a notoriously bad
pick for excited states,\cite{Suellen.Freitas.ea:Cross-Comparisons.2019}
and should never be used over CC3 at the same $\mathcal{O}(N^7)$ 
computational cost. Nevertheless, ADC(3) serves as warning 
that a simple assumption of wavefunction-based methods
outperforming DFT-based methods for excited states has become
obsolete.

The excellent performance observed for the test set of 
Ref.~\citenum{Suellen.Freitas.ea:Cross-Comparisons.2019}
is also retained for other test sets as shown for the Thiel
test set \cite{Schreiber.Silva-Junior.ea:Benchmarks.2008,
Silva-Junior.Sauer.ea:Basis.2010, Silva-Junior.Schreiber.ea:Benchmarks.2010}
in Table~\ref{tab:Thiel}.

\begin{table}[b!]
    \centering
    \caption{Mean signed deviation (MSD) and mean absolute deviation (MAD), 
    root mean square deviation (RMSD) for excitation energies of the Thiel test 
    set.\cite{Schreiber.Silva-Junior.ea:Benchmarks.2008}
    Values for other functionals are taken from 
    Refs.~\citenum{Haasler.Maier.ea:Local.2020}, \citenum{Fuerst.Kaupp:Accurate.2023},
    \citenum{Maier.Bahmann.ea:Validation.2016}, and 
    \citenum{Holzer.Klopper:Communication.2017}.
    Note that we always use the current-dependent and gauge-invariant extension
    of the kinetic energy density. All values are in eV.}
    \label{tab:Thiel}
    \begin{tabular}{
    @{\extracolsep{8pt}}
    l
    S[table-format=-1.3]
    S[table-format=1.3]
    S[table-format=-1.3]
    S[table-format=1.3]}
    \toprule
    & \multicolumn{2}{c}{\text{Singlets}} & \multicolumn{2}{c}
    {\text{Triplets}} \\
    \cmidrule{2-3} \cmidrule{4-5}
    \text{Method}     & \text{MSD} & \text{MAD} & \text{MSD} & \text{MAD}  \\
    \midrule
    PBE               & -0.46 & 0.53 & -0.50 & 0.50 \\
    TPSS              & -0.30 & 0.42 & -0.49 & 0.49 \\
    PBE0              &  0.03 & 0.23 & -0.49 & 0.49 \\lop
    B3LYP             & -0.08 & 0.26 & -0.45 & 0.45 \\
    TPSSh             & -0.12 & 0.29 & -0.49 & 0.49 \\
    CAM-B3LYP         &  0.19 & 0.29 & -0.41 & 0.42 \\
   {\textomega}B97X-D &  0.20 & 0.29 & -0.31 & 0.31 \\
   LC-{\textomega}PBE &  0.40 & 0.40 & -0.50 & 0.55 \\ 
    LH20t             &  0.19 & 0.28 & -0.11 & 0.18 \\
    {\textomega}LH22t &  0.33 & 0.38 & -0.19 & 0.26 \\
    \midrule
    TMHF              &  0.02 & 0.20 & -0.30 & 0.31 \\
    \textbf{CHYF}     &  0.03 & 0.19 & -0.44 & 0.44 \\
    \textbf{CHYF-B95} & -0.05 & 0.18 & -0.23 & 0.25 \\
    \midrule
    ev$GW$-BSE        & -0.02 & 0.16 & -0.56 & 0.56 \\
    ev$GW$-cBSE       &  0.14 & 0.23 & -0.09 & 0.14 \\
    CC2               &  0.14 & 0.17 &  0.17 & 0.18 \\
    \bottomrule
    \end{tabular}
\end{table}

The trend observed previously is continued for the
Thiel test set, with our newly developed functionals exhibiting exceptionally
good performance for singlet excitations. The CHYF model provides
very accurate excitation energies, being comparable to high level CC2 
and $GW$-Bethe--Salpeter equation (BSE) based models. The prediction of
triplet excitations is more dependent on the chosen correlation model, 
with the B95-based model significantly outperforming the PBE-based model.
This is in line with observations from the correlation-kernel augmented
$GW$-BSE model, were a correlation part of DFT is introduced in 
the BSE to specifically improve triplet excitations.\cite{Holzer.Klopper:Communication.2017}
Further, this observation explains the rather good performance
of LH20t and {\textomega}LH22t on triplet excitations, as both
are based on modified B95 correlation functionals.
\cite{Haasler.Maier.ea:Local.2020, Fuerst.Kaupp:Accurate.2023}
Compared to TMHF, CHYF keeps the excellent performance 
for singlet excited states, while especially the modified
B95 correlation is even more successful at predicting
correct triplet excitation energies.

\subsection{NMR Shifts of Organic Compounds}
\label{subsec:nmr}
NMR shieldings and shifts are among the challenging properties for the
DME ansatz as shown previously. \cite{Holzer.Franzke:Local.2022}
Especially TMHF yielded poor results for organic systems. This
behavior drastically changes with the new functionals as outlined
in Table~\ref{tab:nmr}.
The mean absolute errors range from 0.32\,ppm for TMHF to 0.07\,ppm and
0.06\,ppm for CHYF and CHYF-B95, respectively.
That is, the new functionals are a striking improvement over TMHF.
The improvement over TMHF is confirmed for the $^{13}$C NMR shifts.
TMHF performed poorly for these shifts with
a mean absolute error of 10.4\,ppm and a maximum error of 25.9\,ppm.
These are very large errors, especially compared to the top performer
mPSTS with an MAE of 2.7\,ppm and a maximum error of 14.6\,ppm.
CHYF leads to an MAE and maximum error of 3.3\,ppm and 11.6\,ppm, respectively.
CHYF-B95 is again a minor improvement.

Overall, CHYF performs best for the hydrogen shifts and only mPSTS leads
to smaller errors for carbon shifts. The excellent performance of the
CHYF family for NMR is confirmed by further studies on NMR coupling
constants and shieldings in the Supporting Information.
Therefore, the new functionals eliminate the main weakness of TMHF.

\begin{table}
    \centering
    \caption{Mean signed deviation (MSD) and mean absolute deviation (MAD), 
    root mean square deviation (RMSD) for hydrogen and carbon NMR chemical
    shifts relative to CCSD(T) results for the test set of 
    Ref.~\citenum{Flaig.Maurer.ea:Benchmarking.2014}
    Results with other functionals than CHYF and r$^2$SCAN
    taken from Refs.~\citenum{Holzer.Franzke:Local.2022}
    and \citenum{Holzer.Franzke.ea:Assessing.2021}
    Note that we always use the current-dependent and gauge-invariant extension
    of the kinetic energy density. All values are in ppm.}
    \label{tab:nmr}
    \begin{tabular}{
    @{\extracolsep{8pt}}
    l
    S[table-format=-1.2]
    S[table-format=1.2]
    S[table-format=-2.1]
    S[table-format=2.1]}
    \toprule
    & \multicolumn{2}{c}{\text{$^1$H Shifts}} & \multicolumn{2}{c}
    {\text{$^{13}$C Shifts}} \\
    \cmidrule{2-3} \cmidrule{4-5}
    \text{Method}     & \text{MSD} & \text{MAD} & \text{MSD} & \text{MAD}  \\
    \midrule
    KT3               &  0.07 & 0.14 & -2.6 & 4.5 \\
    PBE               &  0.13 & 0.22 &  4.2 & 4.5 \\
    TPSS              &  0.10 & 0.14 &  2.2 & 2.8 \\
    r$^2$SCAN         &  0.19 & 0.21 &  2.3 & 2.7 \\
    PBE0              &  0.13 & 0.17 &  5.8 & 6.1 \\
    B3LYP             &  0.15 & 0.18 &  5.4 & 6.0 \\
    TPSSh             &  0.10 & 0.13 &  3.1 & 3.2 \\
    CAM-B3LYP         &  0.15 & 0.17 &  7.8 & 8.0 \\
   {\textomega}B97X-D &  0.13 & 0.16 &  6.3 & 6.4 \\
   LC-{\textomega}PBE &  0.13 & 0.17 &  8.9 & 9.1 \\
    LH12ct-SsirPW92   &  0.11 & 0.15 &  6.1 & 6.2 \\
    LH14t-calPBE      &  0.08 & 0.10 &  5.5 & 5.6 \\
    LH20t             &  0.06 & 0.10 &  5.9 & 6.0 \\
    mPSTS             &  0.10 & 0.13 &  2.7 & 2.9 \\
    \midrule
    TMHF              &  0.27 & 0.30 &  10.3 & 10.4 \\
    \textbf{CHYF}     &  0.01 & 0.07 &   3.3 & 3.3 \\
    \textbf{CHYF-B95} & -0.01 & 0.06 &   2.9 & 2.9 \\
    \bottomrule
    \end{tabular}
\end{table}

\subsection{Magnetizabilities of Main-Group Systems}
\label{subsec:magnetizabilities}

Finally, we consider magnetizabilities as a further test to check the robustness
of CHYF for magnetic properties. Results for selected functionals,
including the top performers, are listed in Table~\ref{tab:magnetizability}.
Here, CHYF is again one of the best functionals both in terms of mean absolute
deviation and root mean square deviation. Additionally, the maximum
error is comparably small, amounting to only $10 \cdot 10^{-30}$\,J/T$^2$.
Compared to global and range-separated hybrids, this is a remarkable
improvement and it also outperforms many other local hybrids.
Overall, CHYF-B95 performs best with a very small MAD and RMSD of
$2.85 \cdot 10^{-30}$\,J/T$^2$ and $3.68 \cdot 10^{-30}$\,J/T$^2$,
respectively. The maximum error is also below $10 \cdot 10^{-30}$\,J/T$^2$.
For the MAD, slightly smaller errors of $2.25 \cdot 10^{-30}$\,J/T$^2$ were
observed with so-called strong-correlation local hybrids and the
range-separated local hybrid {\textomega}LH22t leads to an MAD of
$3.09 \cdot 10^{-30}$\,J/T$^2$. \cite{Schattenberg.Wodynski.ea:Revisiting.2023}
Therefore, the performance of the two CHYF functionals is
even more remarkable, as their exchange part is simpler and
they show an excellent SCF and grid convergence as demonstrated
in Sec.~\ref{subsec:numerical_behavior}.

Taking together, CHYF and CHYF-B95 perform excellently for
magnetizabilities. Additionally, CHYF and CHYF-B95 show a good
performance for EPR hyperfine coupling constants. Thus, they are
robust and generally applicable functionals for magnetic properties.

\begin{table}[t]
\centering
\caption{Mean signed deviation (MSD), mean absolute deviation (MAD), 
root mean square deviation (RMSD), and maximum deviation
(Max.) for the magnetizability of 27 molecules compared to CCSD(T) data
\cite{Lutnas.Teale.ea:Benchmarking.2009}
as outlined in Ref.~\citenum{Lehtola.Dimitrova.ea:Benchmarking.2021}.
Values for other functionals are taken from Refs.~\citenum{Holzer.Franzke:Local.2022}
and \citenum{Holzer.Franzke.ea:Assessing.2021}.
Note that we always use the current-dependent and gauge-invariant extension
of the kinetic energy density. All values are in units of $10^{-30}$\,J/T$^2$.}
\label{tab:magnetizability}
\begin{tabular}{
@{\extracolsep{8pt}}
l
S[table-format=-1.2]
S[table-format=1.2]
S[table-format=2.2]
S[table-format=1.1]}
\toprule
Method & \text{MSD}    &  \text{MAD}   & \text{RMSD}  & \text{Max.} \\
\midrule
PBE       & 7.09 & 9.15 & 11.68 & 25.55 \\
TPSS      & 7.49 & 7.83 & 10.19 & 24.13 \\
r$^2$SCAN & 3.45 & 5.05 & 7.15 & 19.72 \\
B3LYP     & 4.55 & 5.44 & 7.47 & 18.46 \\
BH{\&}HLYP & 2.17 & 3.13 & 5.10 & 18.16 \\
PBE0      & 5.59 & 5.98 & 8.75 & 23.33 \\
TPSSh     & 7.58 & 7.67 & 11.00 & 33.22 \\
CAM-B3LYP & 2.41 & 3.74 & 5.38 & 14.11 \\
LC-{\textomega}PBE   & 4.15 & 4.96 & 7.32 & 19.03  \\
{\textomega}B97X-D   & 5.94 & 6.27 & 8.68 & 24.48  \\
LH12ct-SsirPW92 & -1.89 & 3.74 & 4.78 & 10.42 \\
LH14t-calPBE & 1.28 & 3.02 & 4.26 & 13.77 \\
LH20t     &  0.45 & 2.47 & 3.73 & 13.61 \\
mPSTS & 6.83 & 6.85 & 9.27 & 25.69 \\
\midrule
TMHF              & 4.94 & 7.12 & 9.53 & 25.45 \\
\textbf{CHYF}     & -1.17 & 3.02 & 3.84 & 10.11 \\
\textbf{CHYF-B95} & -0.30 & 2.85 & 3.68 & 9.63 \\
\bottomrule
\end{tabular}
\end{table}

\section{Extension Towards Multicomponent Density Functional Theory}
\label{subsec:mcdft-ep}

To underline the generality and transferability of our ansatz, 
we note that Eq.~\ref{eq:chyf_corr} can be modified to be 
compatible with general inter-fermion correlation.
Assuming that the inter-fermion correlation is solely dependent 
on the total density, the spin-averaged formulation of the iso-orbital
limit can be applied straightforwardly.
That is, the electron-electron correlation  length is replaced with the
appropriate electron-fermion correlation length
\begin{align}
\label{eq:zdme_fermion}
\underline{z}^{\text{DME}}_{e p} = & \left( |\underline{U}_{e}^{\text{DME}}|^{-1} + |\underline{U}_{p}^{\text{DME}}|^{-1} \right)
\end{align}
Only the parameters to determine 
$\underline{U}_{p}^{\text{DME}}$ are additionally
required. We go forward by shortly demonstrating this for protons.
Naively assuming that $\beta = 79.873$ is serviceable also for protons, 
we re-optimize $\lambda = 0.5922$ by fitting this value to the hydrogen atom. 
During the fitting procedure, both electron and proton are being 
treated as quantum particles. Note that $E_{\text{C}}^{\text{SC}}$
is not needed for the electron-proton correlation and consequently
no interpolation is applied.

Comparing electron-proton correlation energies from this ansatz with 
recently evaluated correlation energies \cite{Holzer.Franzke:Beyond.2024}
from the random phase approximation (RPA) reveals that Eq.~\ref{eq:zdme_fermion}
indeed delivers reasonable electron-proton correlation energies.
Computational settings are the same as in 
Ref.~\citenum{Holzer.Franzke:Beyond.2024}, see also
the Supporting Information. 
Eq.~\ref{eq:zdme_fermion} is evaluated non-selfconsistently
at the respective TMHF+epc17-1 densities.

\begin{figure}[t]
\centering
\includegraphics[width=1.0\linewidth]{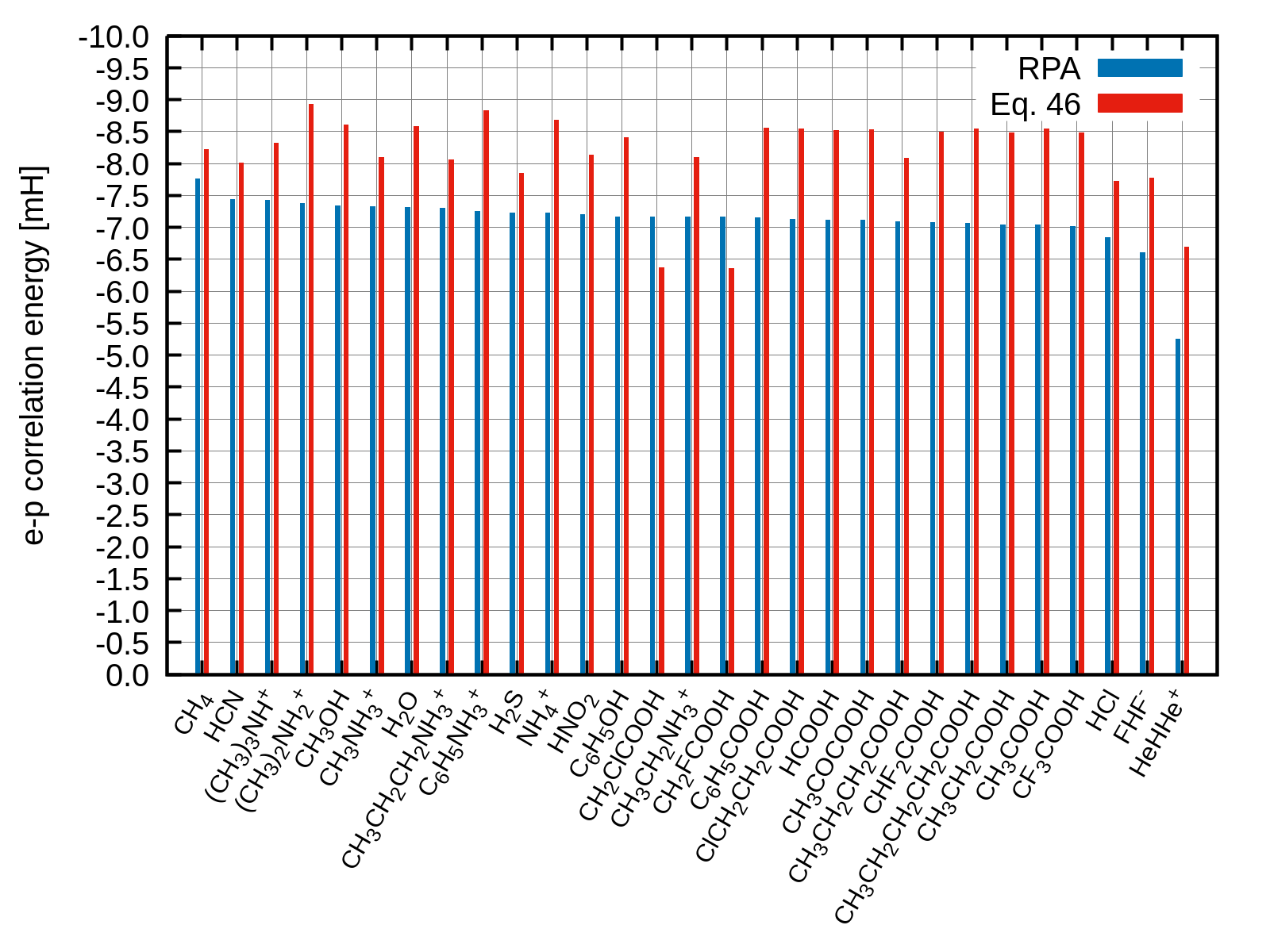}
\caption{Electron-proton correlation energies for 29 molecular
systems obtained at the RPA{\makeatother @}TMHF,
Eq.~\ref{eq:zdme_fermion}{\makeatother @}TMHF+epc17-1, 
and MP2{\makeatother @}HF level
of theory. epc17-1 refers to the electron-proton correlation
functional of Ref.~\citenum{Yang.Brorsen.ea:Development.2017}.
All values are in atomic units (milli-Hartree).}
\label{fig:epcorr}
\end{figure}

As outlined by Fig.~\ref{fig:epcorr}, the electron-proton
correlation energies predicted by Eq.~\ref{eq:zdme_fermion}
are between those obtained from the RPA and those from
MP2. The anomaly at HeHHe$^+$, which 
exhibits an exceptionally low electron-proton correlation
energy is well recovered. Also, FHF$^-$ is correctly
predicted to again have a comparably low electron-proton correlation.
Deficiencies can be seen for the halogenated acetic acid derivatives, 
where Eq.~\ref{eq:zdme_fermion} yields comparably low
correlation energies, while RPA does not find anomalies for
these molecules. The differences are not too high though, 
and especially involving halogen atoms could lead to
more pronounced effects of the neglected self-consistency
in the electron-proton correlation, or hinting at our
quick re-optimization of the parameters $\lambda$ and $\beta$ 
being insufficient.  
While certainly a full re-optimization of all parameters
for the proton would be necessary to yield optimal results, 
we emphasize that the proof of concept of a single
density functional being valid for various different
fermions has been very successful. This is an even more remarkable
result when considering that the local hybrid functional was derived
from first principles by satisfying theoretical constraints.

\section{Conclusion}
\label{sec:conclusion}

In this work, we have derived a local hybrid functional
from theoretical constraints, only taking one- and two-electron
systems into account for exchange and correlation.
Augmenting these with the known gradient expansions 
of the uniform electron gas results in the CHYF functional.
The latter is the first local hybrid 
functional that is fully compatible with a correlation functional
that follows the second-order gradient expansion, yet does 
incorporate large amounts of exact exchange.
CHYF generally exhibits a behavior that resembles an
optimal pure density functionals in many respects
concerning thermochemical properties. Yet, it is strikingly 
different in its ability to predict higher-order properties,
as for example excited states. For the latter, it is shown
that a very accurate description of the excited states can 
be obtained, significantly outperforming any other
density functional. Further investigations of various 
molecular properties of closed-shell and open-shell systems
also outline that our newly developed functional
is robust, leading to clearly acceptable results for all tested
cases. This is a unique feature in density functional theory, 
and can trigger further developments in the direction
of virtually parameter-free density functional approximations.

\begin{suppinfo}
Supporting Information is available with
\begin{itemize}
\item Detailed description of computational methods for all studies
and results for M\"ossbauer isomer shifts, further data
on NMR and EPR properties are presented in the file Supporting-Information.pdf.
\item Spreadsheets with all results are available (W4-11.xlsx, BH76.xlsx,
TDDFT.xlsx, TDDFT-Thiel.xlsx, Mossbauer.xlsx, NMR-Couplings.xlsx,
NMR-Shieldings.xlsx, NMR-Shifts.xlsx, Magnetizabilities.xlsx,
EPR.xlsx, E-P-Correlation.xlsx).
Molecular structures optimized in this work are given in
txt files (Structures-NMR-Couplings.txt).
The uncontracted augmented Dyall-CVTZ basis set (aug-Dyall-CVTZ.txt)
is included.
All of these files are collected in the archive Data.zip.
\item Maple files of the functional are provided in the zip archive
Maple-Files.zip for a straightforward incorporation of the new
functionals into quantum-physical or quantum-chemical software suites.
\end{itemize}
\end{suppinfo}

\section*{Data Availability Statement}
The data that support the findings of this study are available within
the article and its supplementary material.

\section*{Author Contributions and Declarations}
\noindent \textbf{Christof Holzer}: Conceptualization (lead); Data curation (equal);
Formal analysis (equal); Investigation (equal); Methodology (lead);
Software (equal); Validation (equal); Visualization (equal);
Writing – original draft (equal); Writing – review \& editing (equal).
 \\
\noindent \textbf{Yannick J. Franzke}: Conceptualization (supporting); Data curation (equal);
Formal analysis (equal); Investigation (equal); Methodology (supporting);
Software (equal); Validation (equal); Visualization (equal);
Writing – original draft (equal); Writing – review \& editing (equal).
\medskip \newline
\noindent \textbf{Notes} \\
\noindent The authors declare no competing financial interest.

\begin{acknowledgement}
C.H.\ gratefully acknowledges funding by the Volkswagen Foundation.
Y.J.F.\ gratefully acknowledges support via the Walter--Benjamin
programme funded by the Deutsche Forschungsgemeinschaft (DFG, German
Research Foundation) --- 518707327.
\end{acknowledgement}

\bibliography{literature}

\end{document}


\title{Supporting Information: A General and Transferable Local Hybrid Functional for Electronic Structure Theory and Many-Fermion Approaches}
\author{Christof Holzer$^*$}
\affiliation{Institute of Theoretical Solid State Physics, Karlsruhe
Institute of Technology (KIT), Wolfgang-Gaede-Stra\ss{}e 1, 76131 Karlsruhe,
Germany}
\email[Email for correspondence: ]{christof.holzer@kit.edu}

\author{Yannick J. Franzke$^*$}
\affiliation{Otto Schott Institute of Materials Research,
Friedrich Schiller University Jena, L{\"o}bdergraben 32, 07743 Jena, Germany}
\email[Email for correspondence: ]{yannick.franzke@uni-jena.de}

\date{\today}

\maketitle

\tableofcontents

\clearpage
\section{Computational Settings for the Plot of the LMF}
Computational settings for the plot displayed in Fig.~1 of the main
text are as follows.
The local mixing function is calculated for four diatomic molecules at
self-consistent orbitals with the aug-cc-pVQZ \cite{Dunning:Gaussian.1989,
Kendall.Dunning.ea:Electron.1992, Woon.Dunning:Gaussian.1993}
(H, C, O, N, F) or aug-cc-pwCVQZ-DK3 \cite{Bross.Peterson:Correlation.2014}
(Tl) basis sets. Energies were converged with a
threshold of $10^{-8}$\,E$_{\text{h}}$.
The respective bond lengths are 2.1157\,bohr (CO), 2.0494\,bohr (N$_2$),
2.6477\,bohr (F$_2$), and 3.6074\,bohr (TlH).
Large grids \cite{Treutler:Entwicklung.1995, Treutler.Ahlrichs:Efficient.1995}
(grid size 5) are applied for the numerical integration of
the exchange-correlation potential. \cite{Plessow.Weigend:Seminumerical.2012, 
Holzer:improved.2020}
All molecules are aligned along the z-axis, with the center
point of the bond being located at the origin.

\clearpage
\section{Computational Settings for Thermochemistry: W4-11 and BH76 Test Sets}

In line with our previous work, \cite{Holzer.Franzke:Local.2022}
we first study thermochemical properties, i.e.\ atomization energies
and reaction barriers, to test the accuracy of the developed
functionals for the ground-state electronic structure.
For atomization energies, the W4-11 test set is considered~\cite{Karton.Daon.ea:W4-11.2011} 
and the BH76 test is used for the assessment of barrier heights.
\cite{Zhao.Gonzalez-Garcia.ea:Benchmark.2005,
Zhao.Lynch.ea:Multi-coefficient.2005, Goerigk.Grimme:General.2010}
Note that these sets are subsets of the extensive 
``general main group thermochemistry, kinetics, and noncovalent interactions''
set (GMTKN). \cite{Goerigk.Hansen.ea:look.2017} Therefore,
we employ the def2-QZVP basis set~\cite{Weigend.Ahlrichs:Balanced.2005}
and large integration grid (grid size 4) for numerical integration of
the exchange-correlation energy and potential.
\cite{Treutler.Ahlrichs:Efficient.1995, Treutler:Entwicklung.1995}
Self-consistent field (SCF) energies are converged with the default
settings ($10^{-7}$\,E$_{\text{h}}$) of TURBOMOLE.
\cite{Ahlrichs.Bar.ea:Electronic.1989, Furche.Ahlrichs.ea:Turbomole.2014,
Balasubramani.Chen.ea:TURBOMOLE.2020, Franzke.Holzer.ea:TURBOMOLE.2023,
TURBOMOLE, TURBOMOLE-manual}
Results for other functionals are taken from Refs.~\citenum{Holzer.Franzke:Local.2022}
and \citenum{Goerigk.Hansen.ea:look.2017} (all other functionals).

\clearpage
\section{Computational Settings for Excitation Energies from TDDFT}

Besides thermochemistry, excitation energies are of utmost
importance for the applicability of new functionals. Here,
we first consider the benchmark set of Ref.~\citenum{Suellen.Freitas.ea:Cross-Comparisons.2019}
to study the performance for excited states within the
adiabatic approximation. For consistency with previous work,
\cite{Suellen.Freitas.ea:Cross-Comparisons.2019, Holzer.Franzke.ea:Assessing.2021, 
Holzer.Franzke:Local.2022}
the aug-cc-pVTZ basis set \cite{Dunning:Gaussian.1989,
Kendall.Dunning.ea:Electron.1992, Woon.Dunning:Gaussian.1993} 
is employed and the excitation
energies are corrected with the zero-point vibrational energies at
the B3LYP level. \cite{Loos.Jacquemin:Chemically.2019}
Tight SCF thresholds of $10^{-9}$\,E$_{\text{h}}$ and $10^{-7}$\,a.u.\ for
the change of the density matrix are applied, whereas
the response equations are converged with a threshold of $10^{-7}$\,a.u.\ 
for the norm of the residuum.
Large integration grids (grid size 4) are applied.
\cite{Treutler.Ahlrichs:Efficient.1995, Treutler:Entwicklung.1995}

For the Thiel test set, \cite{Schreiber.Silva-Junior.ea:Benchmarks.2008,
Silva-Junior.Sauer.ea:Basis.2010, Silva-Junior.Schreiber.ea:Benchmarks.2010}
settings in TURBOMOLE are chosen as done in Ref.~\citenum{Gui.Holzer.ea:Accuracy.2018}.
That is, SCF calculations are converged with a threshold of
$10^{-8}$\,E$_{\text{h}}$, while TDDFT calculations use a
criterion of $10^{-6}$\,a.u.\ for the norm of the residuum.
\cite{Furche.Krull.ea:Accelerating.2016} Medium-sized grids
(grid size m4) are applied \cite{Treutler:Entwicklung.1995, Treutler.Ahlrichs:Efficient.1995}
and the def2-TZVP basis set is chosen. \cite{Weigend.Ahlrichs:Balanced.2005}
Note that the ground-state and excited-state DFT calculations employ
the resolution of the identity approximation for the Coulomb term
\cite{Eichkorn.Treutler.ea:Auxiliary.1995, Weigend.Kattannek.ea:Approximated.2009,
Bauernschmitt.Haser.ea:Calculation.1997}
(RI-$J$). In both cases tailored auxiliary basis sets are applied.
For the ground-state DFT calculations these are constructed by
fitting the electron density, \cite{Weigend:Accurate.2006}
whereas excited-state calculations use the MP2-fitting basis.
\cite{Weigend.Haser.ea:RI-MP2.1998, Hattig:Optimization.2005}
This leads to a better description of orbital products in the integrals.
\cite{TURBOMOLE-manual, Balasubramani.Chen.ea:TURBOMOLE.2020}

\clearpage
\section{Computational Settings for Multicomponent DFT Calculations}

Mutlicomponent DFT calculations are carried out with the
\texttt{ridft} module utilizing a multicomponent augmented Roothaan--Hall
solver and the resolution of the identity approximation with a common
Hilbert space. \cite{Holzer.Franzke:Beyond.2024}
The def2-QZVPP electronic basis set \cite{Weigend.Ahlrichs:Balanced.2005}
is employed for the non-quantum, i.e.\ classical, nuclei. For
the quantum protons, the def2-QZVPP-mc electronic basis set
\cite{Holzer.Franzke:Beyond.2024} and
the PB5-G protonic basis set \cite{Yu.Pavosevic.ea:Development.2020}
are employed. The def2-QZVPP electronic auxiliary basis set \cite{Weigend:Accurate.2006}
is taken for the classical nuclei. For the quantum protons, the common
auxiliary basis set developed in Ref.~\citenum{Holzer.Franzke:Beyond.2024}
is applied for the multicomponent resolution of the identity approximation.
The latter auxiliary basis set was optimized with an automatic
procedure \cite{Lehtola:Straightforward.2021, Lehtola:Automatic.2023}
as implemented in ERKALE. \cite{Lehtola.Hakala.ea:ERKALE.2012}
The numerical integration use medium grids (grid size 3).
\cite{Treutler:Entwicklung.1995, Treutler.Ahlrichs:Efficient.1995}
Tight thresholds of $10^{-9}$\,E$_{\text{h}}$ for the energy and $10^{-6}$
for the root mean square of the density matrix change are applied.
Structures are taken from Ref.~\citenum{Holzer.Franzke:Beyond.2024}.

\clearpage
\section{M\"ossbauer Isomer Shifts and Contact Density}
\label{sec:mossbauer}

\subsection{Theory}
To probe the density at the nuclei, we calculated the
M\"ossbauer isomer shifts for 12 iron compounds as
outlined in Ref.~\citenum{Zhu.Gao.ea:Mossbauer.2020}.
That is, the contact density $\rho^{\text{c}}$
or the effective contact density $\rho^{\text{e}}$
of the iron center of a compound A is calculated at the
scalar or spin--orbit exact two-component (X2C) level of theory
\cite{Dyall:Interfacing.1997, Kutzelnigg.Liu:Quasirelativistic.2005,
Liu.Peng:Exact.2009, Ilias.Saue:infinite-order.2007}
and the isomer shift is obtained by a linear regression with
respect to the experimental findings according to
\begin{equation}
\delta^{\text{IS}}_{\text{A}} = \alpha \left( \rho_{\text{A}}^{\text{e}} - C \right) + \beta
\end{equation}
with $\alpha$ and $\beta$ denoting fit parameters.
$C$ is kept fixed based on the absolute value of
the (effective) contact density. \cite{Zhu.Gao.ea:Mossbauer.2020, Filatov:First.2009, 
Filatov.Zou.ea:Analytic.2012, Yoshizawa.Filatov.ea:Calculation.2019, 
Hedegard.Knecht.ea:Theoretical.2014} In line with 
Ref.~\citenum{Hedegard.Knecht.ea:Theoretical.2014}, we set
$C$ to 14900 mm/s.
The contact density is computed from the expectation
value of the density operator at the respective nuclei.
At the X2C level, this necessitates the picture-change correction,
\cite{Knecht.Fux.ea:Mossbauer.2011} which we have implemented
by interfacing the density operator into the existing code
infrastructure, similar to the dipole operator in length gauge.
\cite{Kehry.Franzke.ea:Quasirelativistic.2020, Franzke:Calculation.2021}
The effective contact density can be obtained as
the derivative of the energy with respect to the
root-mean-square (RMS) radius of the finite nucleus
model according to
\begin{equation}
\rho_{\text{A}}^{\text{e}} = \left[ \frac{3}{4 \pi Z_{\text{Fe}} ~ \sqrt{ <R_{\text{Fe}}>^2}}
~ \frac{\partial E_{\text{A}}}{\partial \sqrt{ <R_{\text{Fe}}>^2}}
\right]
\label{eq:eff-dens}
\end{equation}
where the finite nuclear radius of the Gaussian charge distribution
can be taken from Ref.~\citenum{Visscher.Dyall:Dirac-Fock.1997} and
$Z_{\text{Fe}} = 26$. That is, only the electron-nucleus potential
and the relativistically modified potential depend on the RMS
radius and consequently many terms for the analytical derivative
of the energy vanish. The respective integral derivative of the potential
can be evaluated straightforwardly as a three-center overlap integral
by considering the derivative of the error function,
c.f.~Refs.~\citenum{Dyall.Fgri-Jr:Finite.1993, Hennum.Klopper.ea:Direct.2001}.
Eq.~\ref{eq:eff-dens} requires to solve first-order X2C response equations.
\cite{Filatov.Zou.ea:Analytic.2012} In the course of the present work,
this approach was also implemented in TURBOMOLE at the scalar X2C and
spin--orbit X2C levels \cite{Peng.Middendorf.ea:efficient.2013,
Franzke.Middendorf.ea:Efficient.2018, Franzke.Weigend:NMR.2019}
and validated against Refs.~\citenum{Filatov.Zou.ea:Analytic.2012, 
Hedegard.Knecht.ea:Theoretical.2014, Yoshizawa.Filatov.ea:Calculation.2019}.
The RMS radius can be evaluated with the
real atomic mass or the isotope number. Herein, we have chosen the
second option for consistency with Ref.~\citenum{Pollak.Weigend:Segmented.2017}.
Furthermore, the transformation to the linear-independent basis
is turned off for the effective contact density in the present work
to improve the numerical accuracy.

\clearpage
\subsection{Computational Details}
Calculations on the Fe compounds
are performed with HF, PBE, \cite{Perdew.Burke.ea:Generalized.1996}
PBE0, \cite{Adamo.Barone:Toward.1999},
r$^2$SCAN, \cite{Furness.Kaplan.ea:Accurate.2020, Furness.Kaplan.ea:Correction.2020}
r$^2$SCANh, \cite{Bursch.Neugebauer.ea:Dispersion.2022}
{\textomega}B97X-D, \cite{Chai.Head-Gordon:Long-range.2008}
LHJ14, \cite{Johnson:Local-hybrid.2014}
LH12ct-SsirPW92, \cite{Arbuznikov.Kaupp:Importance.2012}
LH14t-calPBE, \cite{Arbuznikov.Kaupp:Towards.2014}
LH20t, \cite{Haasler.Maier.ea:Local.2020}
Tao-Mo (TM), \cite{Tao.Mo:Accurate.2016}
TMHF, \cite{Holzer.Franzke:Local.2022}
and CHYF,
as well as the post-Kohn--Sham random phase approximation (RPA)
based on PBE orbitals. \cite{Eshuis.Yarkony.ea:Fast.2010, Eshuis.Bates.ea:Electron.2012,
Burow.Bates.ea:Analytical.2014, Chen.Voora.ea:Random.2017}
In RPA, the (effective) contact density is obtained as an expectation value with the
relaxed density, see also Ref.~\citenum{Bruder.Weigend.ea:Application.2024}.
For comparison, we also included a global hybrid version of the TM functional
with 10\% of exact exchange (TMh).
We use Libxc \cite{Marques.Oliveira.ea:Libxc.2012, Lehtola.Steigemann.ea:Recent.2018, 
LIBXC.2023} for r$^2$SCAN, r$^2$SCANh, TM, TMh, and {\textomega}B97X-D.
The scalar X2C approach is applied, as spin--orbit effects are shown
to be negligible for the isomer shift, as already observed previously
in Ref.~\citenum{Hedegard.Knecht.ea:Theoretical.2014}.
The x2c-QZVPall orbital basis set is employed for all elements except
for Fe. \cite{Franzke.Spiske.ea:Segmented.2020} Here, we apply the
uncontracted Dyall-CVTZ basis sets \cite{Dyall:Relativistic.2006, Dyall-Repo}
of the Dirac program \cite{Saue.Bast.ea:DIRAC.2020}
augmented by steep $s$ and $p$ functions. \cite{Hedegard.Knecht.ea:Theoretical.2014}
The x2c-QZVPall auxiliary basis sets (jbas) is used for all atoms
\cite{Pollak.Weigend:Segmented.2017, Franzke.Spiske.ea:Segmented.2020}
with the resolution of the identity approximation to the Coulomb integrals
\cite{Eichkorn.Treutler.ea:Auxiliary.1995, Weigend.Kattannek.ea:Approximated.2009}
(RI-$J$).
Comparisons to a large even-tempered basis \cite{Franzke.Spiske.ea:Segmented.2020}
show that this is sufficient, whereas application of the
x2c-QZVPall bases for all atoms clearly underestimates the
absolute values of the contact densities. Energies are converged
with a threshold of $10^{-8}$\,E$_{\text{h}}$ and large grids
(grid size 4a) are employed. \cite{Franzke.Tress.ea:Error-consistent.2019}
The 2c calculations use a threshold of $10^{-7}$\,E$_{\text{h}}$.
To account for the counter ions, the conductor-like screening model (COSMO)
is applied with the default setting ($\epsilon_{\text{r}} = \infty$).
\cite{Klamt.Schuurmann:COSMO.1993, Schafer.Klamt.ea:COSMO.2000}
We note in passing that these are the first calculations of
effective contact densities and M\"ossbauer isomer shifts with
current-dependent functionals and local hybrids, at least
to the best of our knowledge.
For the RPA calculations, COSMO is turned off as it is not yet implemented
for the Z-vector equations to calculate the relaxed density. 
\cite{Burow.Bates.ea:Analytical.2014} For consistency, the PBE orbitals
for RPA were also calculated without COSMO, except for FeF$_6^{4-}$.
Here, neglecting COSMO leads to serious spin contamination and the RPA results
are clearly worsened. We use the Gauss--Legendre method with 150 points for the
imaginary frequency calculation. Convergence of the integration with respect
to the number of points was confirmed by increasing it to 200, i.e.\ the
RPA correlation energy changes by less than $10^{-5}$\,E$_{\text{h}}$. In contrast,
the RPA correlation energies with the Clenshaw--Curtis method are not converged
up to $10^{-4}$\,E$_{\text{h}}$ even with 200 points. For instance, the
correlation energy of FeCl$_4^-$ changes by $-6 \cdot 10^{-4}$\,E$_{\text{h}}$
when increasing the number of points from 150 to 200 and by
$-3 \cdot 10^{-4}$\,E$_{\text{h}}$ upon an increase from 200 to 300 points.
Structures are taken from Refs.~\citenum{Romelt.Ye.ea:Calibration.2009,
Neese:Prediction.2002, Hopmann.Ghosh.ea:Density.2009}.
Experimental results were collected in Ref.~\citenum{Zhu.Gao.ea:Mossbauer.2020}.

\clearpage
\subsection{Results}

The M\"ossbauer isomer shifts for iron compounds are listed in
Tab.~\ref{tab:mossbauer}. The contact densities and effective
contact densities are given in the Supporting Information (Mossbauer.xlsx).
For CHYF, results from scalar-relativistic one-component (1c) and
two-component (2c) calculations are further compared in
Tab.~\ref{tab:mossbauer-so}. The 2c generalization is available
with a current-dependent generalization of the kinetic energy
density (cCHYF). \cite{Holzer.Franke.ea:Current.2022} This
shows that the scalar-relativistic approximation is sufficient
and the impact of spin--orbit coupling is negligible for
the iron compounds.

Furthermore, The effective contact densities are approximately 1\%
smaller than the contact densities. For all functionals,
a constant ratio of $\rho^{\text{c}}/\rho^{\text{e}}$ is found.
That is, the linear regression is valid for both densities
and the coefficient of determination $R^2$ is better than 0.98
for all density functional approximations and RPA.

Considering the results in Tab.~\ref{tab:mossbauer},
all density functional approximations applied herein
perform well compared to the experimental findings.
For FeBr$_4^-$ and FeCl$_4^-$, very similar (effective) contact densities
are observed. The same holds for the isomer shift and a different
trend of isomer shift is found than in the experiment. Note that
this was also observed previously with wavefunction-based methods.
\cite{Zhu.Gao.ea:Mossbauer.2020}
The largest absolute errors are found for FeS$_8$C$_8$O$_4^{2-}$ and FeCl$_4^-$
with almost 0.1\,mm/s for most functionals. The RMSE of wavefunction-based
methods such as spin component-scaled second-order M\"oller--Plesset
perturbation theory (SCS-MP2) and the iterative configuration expansion
(ICE) self-consistent field (SCF) method is $6.1 \cdot 10^{-2}$ and
$8.0 \cdot 10^{-2}$\,mm/s, respectively. \cite{Zhu.Gao.ea:Mossbauer.2020}
Therefore, DFT outperforms SCS-MP2 and ICE-SCF for the iron compounds
herein.

Second, the admixture of exact exchange generally leads to systematically
better results in terms of quantitative agreement, i.e.\ a smaller RSMD
and MAE values. RPA leads to excellent results and clearly outperforms
semilocal functionals and most hybrids. PBE0 is also one of the top performers
and reduces the RMSE of PBE. Overall, hybrid functionals clearly outperform their
semilocal counterparts. Local hybrids are no consistent improvement upon
PBE0, however, they yield very good results. CHYF leads to a similar
MAE and RMSE as TMHF. Overall, LH12ct-SsirPW92 is the top performer
with the smallest MAE and RMSE of all functionals.

Third, the nuclear calibration constant {\textalpha} of all functionals
is in excellent agreement with the literature.
\cite{Sinnecker.Slep.ea:Performance.2005, Kurian.Filatov:DFT.2008, 
Hedegard.Knecht.ea:Theoretical.2014, Gubler.Finkelmann.ea:Theoretical.2013}
Other relativistic DFT methods yielded a constant between $-0.26$ and
$-0.29$\,bohr$^3$\,mm/s. For the same functional family, the admixture
of exact exchange leads to a decrease of the nuclear calibration constant.

\clearpage
\begin{sidewaystable}[t!]
\centering
\footnotesize
\caption{M{\"o}ssbauer isomer shifts $\delta^{\text{IS}}$ in mm/s
for 12 iron compounds as well as the mean absolute error (MAE) and
the root-mean-square  error (RMSE) both in units of $10^{-2}$\,mm/s.
Experimental references (Expt.) were collected in Ref.~\citenum{Zhu.Gao.ea:Mossbauer.2020}.
The augmented Dyall-CVTZ (Fe) and x2c-QZVPall (other)
basis sets are applied. The fit parameters $\alpha$ and $\beta$ are given in
bohr$^3$\,mm/s and bohr$^{-3}$, respectively. For all functionals and RPA,
the coefficient of determination $R^2$ is better than 0.98.
LH12ct-SSirPW92 and LH14t-calPBE are shortened to LH12ct and LH14t in this table.}
\label{tab:mossbauer}
\begin{tabular}{
@{\extracolsep{1pt}}
l
S[table-format = -2.2]
S[table-format = -2.2]
S[table-format = -2.2]
S[table-format = -2.2]
S[table-format = -2.2]
S[table-format = -2.2]
S[table-format = -2.2]
S[table-format = -2.2]
S[table-format = -2.2]
S[table-format = -2.2]
S[table-format = -2.2]
S[table-format = -2.2]
S[table-format = -2.2]
S[table-format = -2.2]
S[table-format = -2.2]
S[table-format = -2.2]
S[table-format = -2.2]
}
\toprule
Compound & {\text{HF}} & {\text{PBE}} & {\text{PBE0}} & {\text{r$^2$SCAN}} &
{\text{r$^2$SCANh}} & {\text{{\textomega}B97X-D}} & {\text{LHJ14}} &
{\text{LH12ct}} & {\text{LH14t}} & {\text{LH20t}} & {\text{TM}} & 
{\text{TMh}} & {\text{TMHF}} & {\text{CHYF}} & {\text{CHYF-B95}} &  {\text{RPA}} & {\text{Expt.}} \\
\midrule
FeBr$_4^-$ & -0.02 & 0.32 & 0.24 & 0.30 & 0.27 & 0.24 & 0.30 & 0.26 & 0.26 & 0.25 & 0.32 & 0.29 & 0.28 & 0.30 & 0.30 & 0.28 & 0.25 \\
FeCl$_4^-$ & 0.05 & 0.33 & 0.26 & 0.32 & 0.29 & 0.26 & 0.31 & 0.28 & 0.28 & 0.27 & 0.34 & 0.31 & 0.29 & 0.31 & 0.31 & 0.28 & 0.19 \\
FeCl$_4^{2-}$ & 1.03 & 0.82 & 0.87 & 0.85 & 0.87 & 0.87 & 0.84 & 0.86 & 0.85 & 0.86 & 0.83 & 0.85 & 0.86 & 0.85 & 0.84 & 0.87 & 0.90 \\
Fe(CN)$_6^{3-}$ & -0.22 & -0.13 & -0.08 & -0.12 & -0.10 & -0.07 & -0.10 & -0.09 & -0.09 & -0.08 & -0.13 & -0.11 & -0.10 & -0.13 & -0.12 & -0.13 & -0.13 \\
Fe(CN)$_6^{4-}$ & 0.10 & -0.07 & 0.00 & -0.06 & -0.03 & 0.03 & -0.01 & -0.02 & -0.01 & 0.01 & -0.07 & -0.04 & 0.01 & -0.04 & -0.03 & -0.06 & -0.02 \\
Fe(CO)$_5$ & -0.17 & -0.24 & -0.16 & -0.24 & -0.21 & -0.13 & -0.17 & -0.18 & -0.17 & -0.15 & -0.24 & -0.21 & -0.15 & -0.18 & -0.17 & -0.19 & -0.18 \\
FeF$_6^{3-}$ & 0.49 & 0.58 & 0.53 & 0.57 & 0.55 & 0.53 & 0.54 & 0.55 & 0.54 & 0.54 & 0.57 & 0.55 & 0.53 & 0.55 & 0.55 & 0.49 & 0.48 \\
FeF$_6^{4-}$ & 1.43 & 1.27 & 1.31 & 1.29 & 1.30 & 1.28 & 1.27 & 1.31 & 1.31 & 1.31 & 1.25 & 1.27 & 1.27 & 1.27 & 1.27 & 1.33 & 1.34 \\
Fe(H$_2$O$_5$)NO$^{2+}$ & 0.44 & 0.82 & 0.79 & 0.81 & 0.79 & 0.79 & 0.82 & 0.80 & 0.80 & 0.79 & 0.82 & 0.81 & 0.80 & 0.81 & 0.82 & 0.86 & 0.76 \\
Fe(H$_2$O$_6$)$^{3+}$ & 0.43 & 0.58 & 0.50 & 0.55 & 0.52 & 0.49 & 0.52 & 0.51 & 0.51 & 0.49 & 0.57 & 0.54 & 0.49 & 0.52 & 0.53 & 0.48 & 0.51 \\
FeO$_4^{2-}$ & -0.41 & -0.88 & -0.94 & -0.88 & -0.90 & -0.98 & -0.95 & -0.93 & -0.93 & -0.95 & -0.89 & -0.91 & -0.96 & -0.92 & -0.94 & -0.86 & -0.87 \\
FeS$_8$C$_8$O$_4^{2-}$ & 0.76 & 0.51 & 0.57 & 0.52 & 0.55 & 0.58 & 0.54 & 0.56 & 0.55 & 0.57 & 0.53 & 0.55 & 0.58 & 0.55 & 0.54 & 0.55 & 0.67 \\
\midrule
$\alpha$ & -0.26 & -0.31 & -0.27 & -0.29 & -0.28 & -0.27 & -0.30 & -0.26 & -0.27 & -0.26 & -0.30 & -0.28 & -0.28 & -0.28 & -0.28 & -0.28 & \\
$\beta$ & -2.87 & 10.52 & 6.25 & -1.17 & -1.33 & 25.18 & 16.23 & -10.43 & -20.45 & -1.58 & 10.14 & 8.45 & 25.40 & -12.11 & -12.42 & -3.15 & \\
$R^2$ & 0.88 & 0.98 & 0.99 & 0.98 & 0.99 & 0.99 & 0.99 & 0.99 & 0.99 & 0.99 & 0.98 & 0.99 & 0.99 & 0.99 & 0.99 & 0.99 & \\
MAE & 15.08 & 7.30 & 4.17 & 5.99 & 4.51 & 5.20 & 5.62 & 4.02 & 4.26 & 4.54 & 7.29 & 5.61 & 5.09 & 5.13 & 5.65 & 3.99 & \\
RSME & 19.79 & 8.58 & 4.95 & 7.26 & 5.64 & 5.85 & 6.75 & 5.21 & 5.43 & 5.30 & 8.34 & 6.51 & 5.84 & 6.40 & 6.85 & 5.59 & \\
\bottomrule
\end{tabular}
\end{sidewaystable}

\clearpage
\begin{table}[t!]
\centering
\caption{M{\"o}ssbauer isomer shifts $\delta^{\text{IS}}$ in mm/s
for 12 iron compounds as well as the mean absolute error (MAE) and
the root-mean-square error (RMSE) both in units of $10^{-2}$\,mm/s.
Comparison of scalar-relativistic one-component (1c) and
two-component (2c) calculations with CHYF. The 2c generalization
is available with a current-dependent generalization of the
kinetic energy density (cCHYF).
Experimental references (Expt.) were collected in 
Ref.~\citenum{Zhu.Gao.ea:Mossbauer.2020}.}
\label{tab:mossbauer-so}
\begin{tabular}{
@{\extracolsep{32pt}}
l
S[table-format = -2.2]
S[table-format = -2.2]
S[table-format = -2.2]
S[table-format = -2.2]
@{}
}
\toprule
Compound & {\text{1c CHYF}} & {\text{2c CHYF}} & {\text{2c cCHYF}} & {\text{Expt.}} \\
\midrule
FeBr$_4^-$ & 0.30 & 0.30 & 0.30 & 0.25 \\
FeCl$_4^-$ & 0.31 & 0.31 & 0.31 & 0.19 \\
FeCl$_4^{2-}$ & 0.85 & 0.85 & 0.85 & 0.90 \\
Fe(CN)$_6^{3-}$ & -0.13 & -0.13 & -0.13 & -0.13 \\
Fe(CN)$_6^{4-}$ & -0.04 & -0.04 & -0.04 & -0.02 \\
Fe(CO)$_5$ & -0.18 & -0.18 & -0.18 & -0.18 \\
FeF$6^{3-}$ & 0.55 & 0.55 & 0.55 & 0.48 \\
FeF$_6^{4-}$ & 1.27 & 1.27 & 1.27 & 1.34 \\
Fe(H$_2$O$_5$)NO$^{2+}$ & 0.81 & 0.81 & 0.81 & 0.76 \\
Fe(H$_2$O$_6$)$^{3+}$ & 0.52 & 0.52 & 0.52 & 0.51 \\
FeO$_4^{2-}$ & -0.92 & -0.92 & -0.92 & -0.87 \\
FeS$_8$C$_8$O$_4^{2-}$ & 0.55 & 0.55 & 0.55 & 0.67 \\
\midrule
$\alpha$ & -0.28 & -0.28 & -0.28 & \\
$\beta$ & -12.11 & -11.91 & -11.91 & \\
$R^2$ & 0.99 & 0.99 & 0.99 & \\
MAE & 5.13 & 5.15 & 5.15 & \\
RSME & 6.40 & 6.42 & 6.41 & \\
\bottomrule
\end{tabular}
\end{table}

\clearpage
\section{NMR Indirect Spin--Spin Coupling Constants of Main-Group Systems}
\label{sec:sscc}

\subsection{Computational Details}
Furthermore, nuclear magnetic resonance (NMR) coupling constants are calculated
for the test set of Ref.~\citenum{Faber.Sauer.ea:Importance.2017} using the DFT
protocol of Ref.~\citenum{Holzer.Franzke.ea:Assessing.2021}, which is based
on Ref.~\citenum{Mack.Schattenberg.ea:Nuclear.2020}. Therefore,
structures are optimized with the aug-cc-pVQZ basis sets, \cite{Dunning:Gaussian.1989,
Kendall.Dunning.ea:Electron.1992, Woon.Dunning:Gaussian.1995, CC-Repo} while
NMR coupling constants are computed with the aug-ccJ-pVTZ basis set
\cite{Benedikt.Auer.ea:Optimization.2008} as taken from the Basis Set Exchange
library. \cite{Feller:role.1996, Schuchardt.Didier.ea:Basis.2007,
Pritchard.Altarawy.ea:New.2019, BSE-library}
Large integration grids (grid size 4) are applied for the DFT part
\cite{Treutler.Ahlrichs:Efficient.1995, Treutler:Entwicklung.1995} and
tight SCF thresholds of $10^{-9}$\,E$_{\text{h}}$ and $10^{-9}$\,a.u.\
for the norm of the density matrix changes are chosen. Response equations
are converged with a threshold of $10^{-9}$\,a.u.\ for the norm of the residuum.
\cite{Furche.Krull.ea:Accelerating.2016} Inclusion of the current density
\cite{Bates.Furche:Harnessing.2012, Holzer.Franzke.ea:Assessing.2021}
is denoted by a ``c'' in the functional acronym (e.g.\ cTPSS) for all
magnetic properties. We consider the  
KT3, \cite{Keal.Tozer:semiempirical.2004}
BP86, \cite{Perdew:Density-functional.1986, Becke:Density-functional.1988}
PBE, \cite{Perdew.Burke.ea:Generalized.1996}
TPSS, \cite{Tao.Perdew.ea:Climbing.2003}
r$^2$SCAN, \cite{Furness.Kaplan.ea:Accurate.2020,Furness.Kaplan.ea:Correction.2020}
Tao-Mo (TM), \cite{Tao.Mo:Accurate.2016}
BH{\&}HLYP, \cite{Becke:Density-functional.1988, Lee.Yang.ea:Development.1988, Becke:new.1993}
B3LYP, \cite{Lee.Yang.ea:Development.1988,
Becke:Density-functional.1993, Stephens.Devlin.ea:Ab.1994}
PBE0, \cite{Perdew.Burke.ea:Generalized.1996, Adamo.Barone:Toward.1999}
TPSSh, \cite{Staroverov.Scuseria.ea:Comparative.2003}
TPSS0, \cite{Staroverov.Scuseria.ea:Comparative.2003, Grimme:Accurate.2005}
r$^2$SCANh, \cite{Furness.Kaplan.ea:Accurate.2020, Furness.Kaplan.ea:Correction.2020, Bursch.Neugebauer.ea:Dispersion.2022}
r$^2$SCAN0, \cite{Furness.Kaplan.ea:Accurate.2020, Furness.Kaplan.ea:Correction.2020, Bursch.Neugebauer.ea:Dispersion.2022}
r$^2$SCAN50, \cite{Furness.Kaplan.ea:Accurate.2020, Furness.Kaplan.ea:Correction.2020, Bursch.Neugebauer.ea:Dispersion.2022}
CAM-B3LYP, \cite{Yanai.Tew.ea:new.2004}
CAM-QPT-00, \cite{Jin.Bartlett:QTP.2016}
CAM-QTP-02, \cite{Haiduke.Bartlett:Communication.2018}
HSE06, \cite{Heyd.Scuseria.ea:Hybrid.2003, Heyd.Scuseria.ea:Erratum.2006,
Krukau.Vydrov.ea:Influence.2006}
LC-{\textomega}PBE, \cite{Vydrov.Scuseria:Assessment.2006}
{\textomega}B97X-D, \cite{Chai.Head-Gordon:Long-range.2008}
LH07t-SVWN,\cite{Bahmann.Rodenberg.ea:thermochemically.2007}
LH12ct-SsirPW92, \cite{Arbuznikov.Kaupp:Importance.2012}
LH14t-calPBE, \cite{Arbuznikov.Kaupp:Towards.2014}
LH20t, \cite{Haasler.Maier.ea:Local.2020}
LH20t$^*$ (LH20t without calibration), \cite{Haasler.Maier.ea:Local.2020}
LHJ14, \cite{Johnson:Local-hybrid.2014}
mPSTS, \cite{Perdew.Staroverov.ea:Density.2008, Holzer.Franzke.ea:Assessing.2021}
LHJ-HF, \cite{Holzer.Franzke:Local.2022}
LHJ-HFcal, \cite{Holzer.Franzke:Local.2022}
TMHF, \cite{Holzer.Franzke:Local.2022}
and TMHF-3P \cite{Holzer.Franzke:Local.2022}
density functional approximations for comparison.
Results for these functionals are taken from Refs.~\citenum{Holzer.Franzke:Local.2022,
Holzer.Franzke.ea:Assessing.2021, Franzke.Holzer.ea:NMR.2022}.
Coupling constants with an absolute value below 6\,Hz are neglected
in the statistical evaluation.

\clearpage
\subsection{Results}
\begin{sidewaysfigure}
\includegraphics[width=1.0\linewidth]{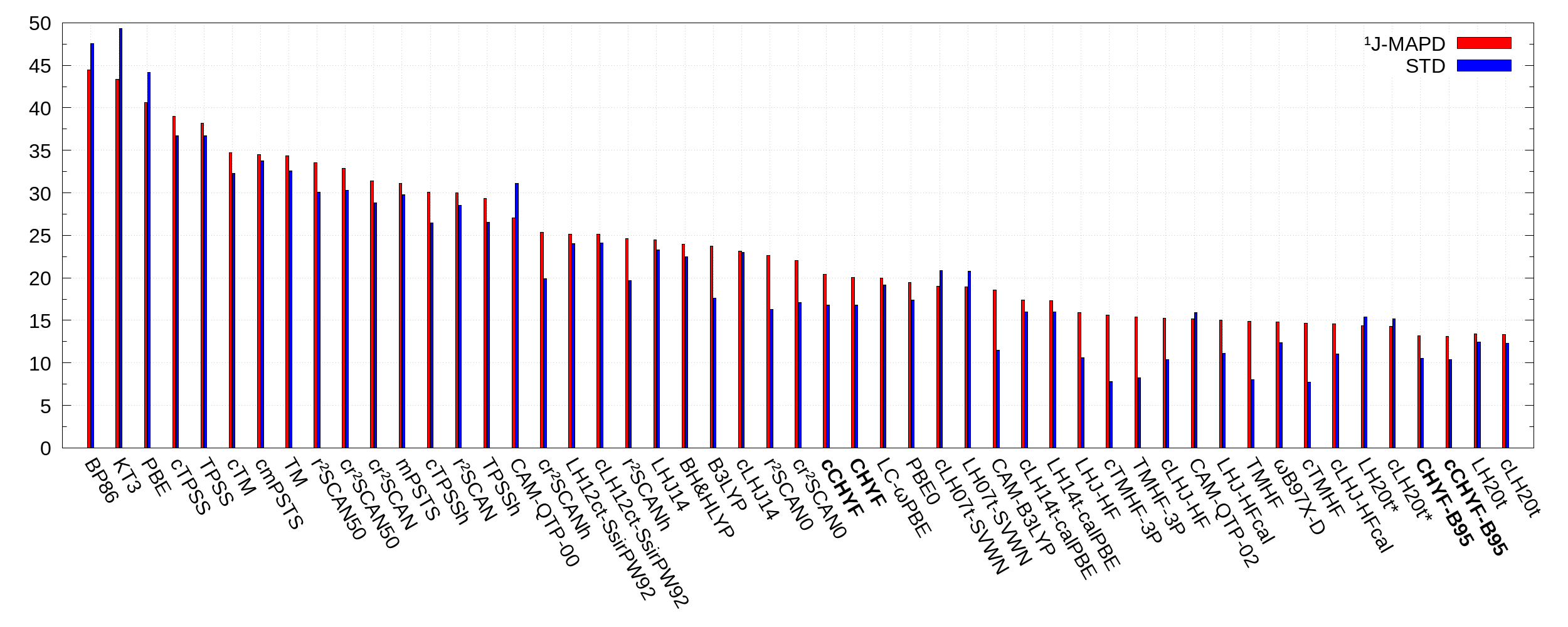}
\caption{Assessment of various density functional approximations
for $^1J$ couplings compared to CC3 results for 13 organic compounds.
\cite{Faber.Sauer.ea:Importance.2017}. A mean absolute percent-wise
error (MAPD) and the standard deviation (STD) are used to statistically
evaluate the performance. Functionals are sorted
according to the mean absolute percent-wise deviation.}
\label{fig:ssccs-1}
\end{sidewaysfigure}

\begin{sidewaysfigure}
\includegraphics[width=1.0\linewidth]{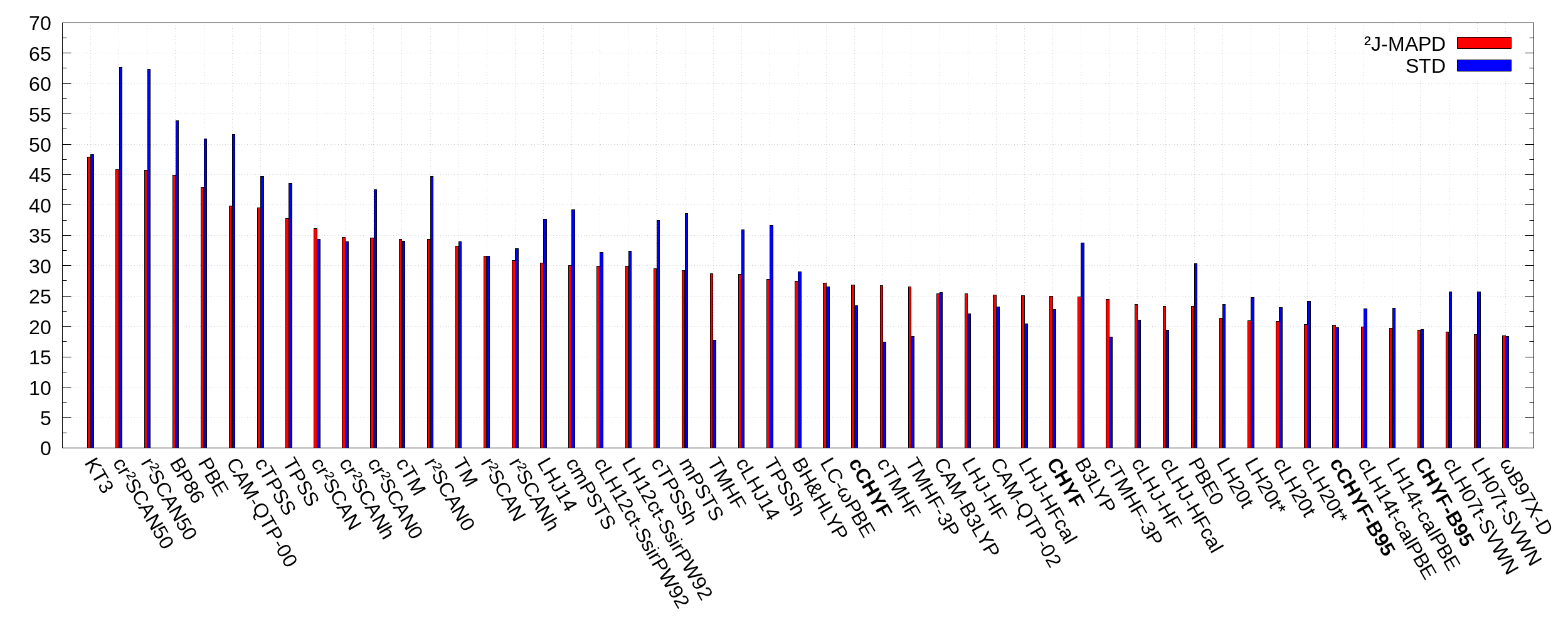}
\caption{Assessment of various density functional approximations
for $^2J$ couplings compared to CC3 results for 13 organic compounds.
\cite{Faber.Sauer.ea:Importance.2017} A mean absolute percent-wise
error (MAPD) and the standard deviation (STD) are used to statistically
evaluate the performance. Functionals are sorted
according to the mean absolute percent-wise deviation.}
\label{fig:ssccs-2}
\end{sidewaysfigure}

Results are illustrated in Figs.~\ref{fig:ssccs-1} and \ref{fig:ssccs-2}
For the $^1J$ coupling constants, the new functional CHYF-B95 is the top performers
with mean percent-wise deviations (MAPDs) of around 13\%. Other functionals such as
LH20t, TMHF, or {\textomega}B97X-D yield a MAPD of around 13.5
to 15\%. The smallest standard deviation is found for TMHF with about 8\%.
The worst functionals in this regard, namely KT3 and BP86, result in an
MAPD of more than 40\% with a standard deviation of almost 50\%.
CHYF leads to an MAPD of 20\% and the standard deviation amounts to 17\%.
These are very similar results as for the well established functionals B3LYP
or BH{\&}HLYP.

Similar findings hold for the $^{2/3}J$ coupling constant. The MAPD
ranges from 18\% for {\textomega}B97X-D to 48\% for KT3. Most
functionals yield errors between 20\% and 30\%. Here, CHYF leads to
an error of about 23\%, which is better than the errors observed
with B3LYP and BH{\&}HLYP.
Notably, none of the r$^2$SCAN hybrids outperforms CHYF.
As for the $^1J$ couplings, CHYF-B95 is among the top performers
for the $^{2/3}J$ coupling constant with an MAPD of less than 20\%.

Overall, CHYF-B95 is the top performer and CHYF performs reasonably well.
This confirms the findings of the excitation energies of the Thiel tests
set, which is well rationalized by the relationship of NMR couplings
to triplet excitations. \cite{Helgaker.Jaszunski.ea:Ab.1999}

\clearpage

\clearpage

\clearpage
\section{NMR Shielding Constants of Main-Group Systems}
\label{sec:shieldings}

\subsection{Computational Details}
NMR shielding constants are calculated for the test set of 
Ref.~\citenum{Stoychev.Auer.ea:Self-Consistent.2018}.
Thus, the large pcSseg-4 basis set is employed. \cite{Jensen:Segmented.2015}
Thresholds and considered functionals are the same as for the
calculation of the NMR couplings. See also Refs.~\citenum{Holzer.Franzke:Local.2022}
and \citenum{Holzer.Franzke.ea:Assessing.2021}. For comparison, we
also included a global hybrid version of the TM functional with 10\% of
exact exchange (TMh). Results for all functionals except for
CHYF, CHYF-B95, and TMh are taken from Refs.~\citenum{Holzer.Franzke:Local.2022}
and \citenum{Holzer.Franzke.ea:Assessing.2021}.
The response equations for the NMR shielding calculations
\cite{Haser.Ahlrichs.ea:Direct.1992, Mas-Huniar:1999, Reiter.Mack.ea:Calculation.2018,
Schattenberg.Reiter.ea:Efficient.2020, Holzer.Franzke.ea:Assessing.2021, 
Schattenberg.Kaupp:Effect.2021, Schattenberg.Kaupp:Implementation.2021,
Holzer:improved.2020} are converged with a threshold of $10^{-7}$\,a.u.\ for the
norm of the residuum. \cite{Furche.Krull.ea:Accelerating.2016}
Functionals depending on the kinetic energy density are generalized with
the vector potential by default. \cite{Maximoff.Scuseria:Nuclear.2004}
This ensures gauge-origin invariance but violates the iso-orbital
constraint. \cite{Bates.Furche:Harnessing.2012}
Inclusion of the current density \cite{Schattenberg.Kaupp:Effect.2021, 
Holzer.Franzke.ea:Assessing.2021} is denoted by a ``c'' in the functional
acronym (e.g.\ cTPSS).

In line with previous work, \cite{Holzer.Franzke.ea:Assessing.2021,
Holzer.Franzke:Local.2022} we statistically evaluate the results for
the hydrogen, carbon, and the other nuclei. To do so,
the mean absolute error (MAE), the mean signed error (MSE),
and its standard deviation (STD) are considered.

\clearpage
\subsection{Results}
\begin{sidewaysfigure}
\includegraphics[width=1.0\linewidth]{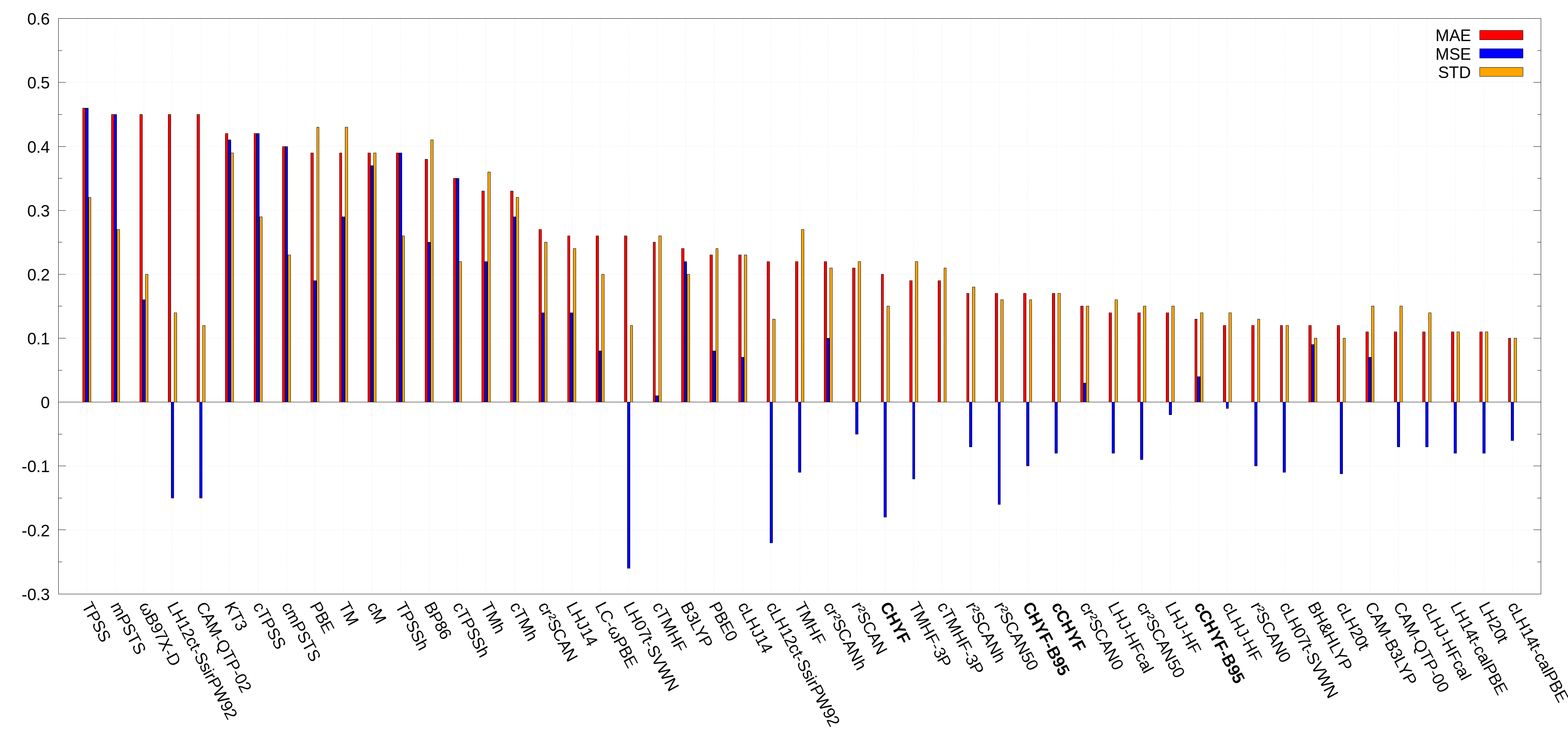}
\caption{Assessment of various density functional approximations
compared to CCSD(T) results ~\cite{Stoychev.Auer.ea:Self-Consistent.2018}
for $^{1}$H NMR shielding constants.
MAE, MSE, and STD denote the mean absolute error, mean signed error, and its standard
deviation.}
\label{fig:1h-shieldings}
\end{sidewaysfigure}

For the $^1$H shieldings in Fig.~\ref{fig:1h-shieldings}, the mean
absolute error ranges from 0.46\,ppm for TPSS to
0.10\,ppm for cLH14t-calPBE. CHYF and TMHF lead to errors of
0.20\,ppm and 0.22\,ppm, using the current-dependent form
slightly increases the errors for TMHF to 0.25\,ppm but decreases
the errors of CHYF to 0.17\,ppm. CHYF-B95 leads to mean errors
of 0.17 and 0.13\,ppm, respectively.
This marks a robust performance for both new functionals.

\begin{sidewaysfigure}
\includegraphics[width=1.0\linewidth]{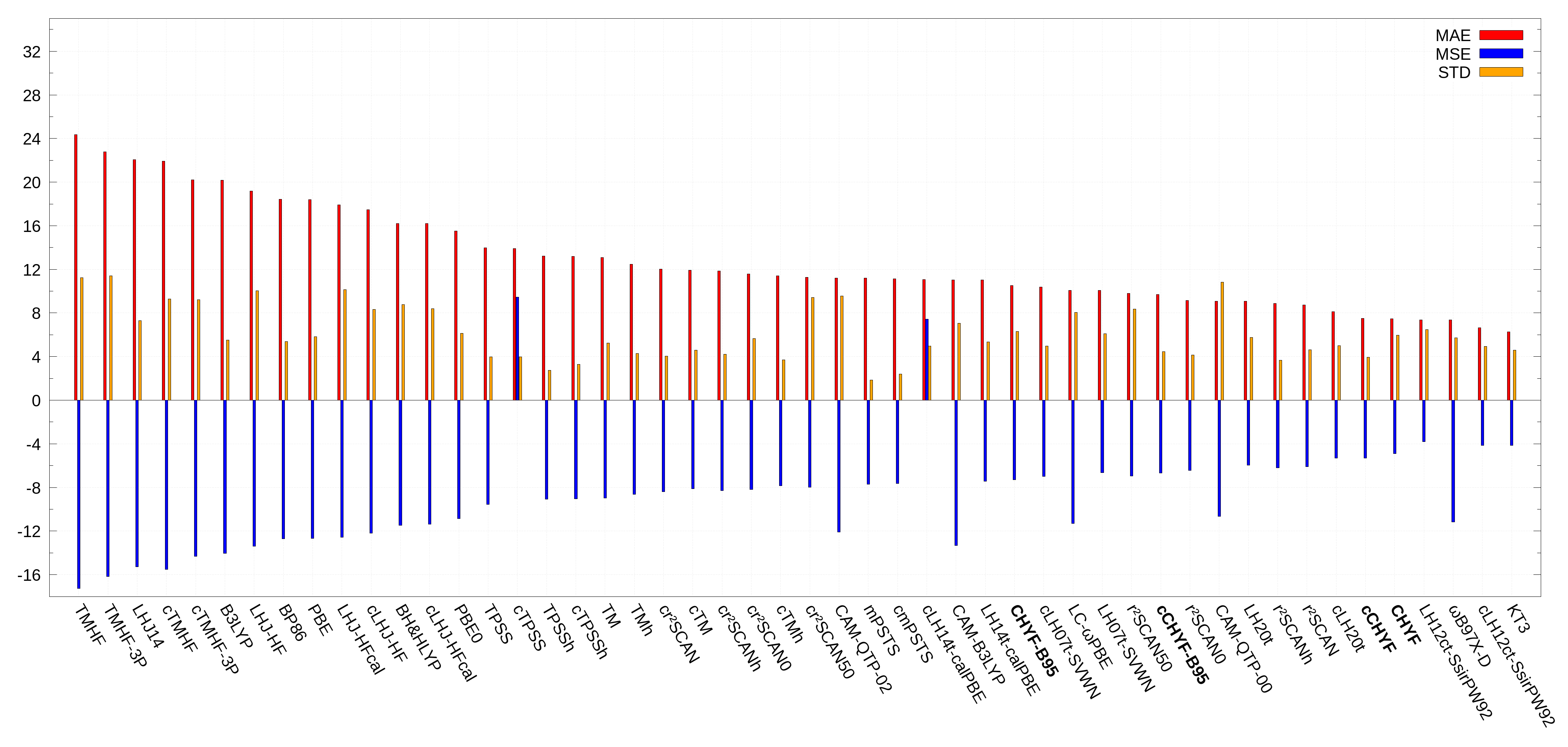}
\caption{Assessment of various density functional approximations
compared to CCSD(T) results ~\cite{Stoychev.Auer.ea:Self-Consistent.2018}
for $^{13}$C NMR shielding constants.
MAE, MSE, and STD denote the mean absolute error, mean signed error, and its standard
deviation.}
\label{fig:13c-shieldings}
\end{sidewaysfigure}

The $^{13}$C shieldings in Fig.\ref{fig:13c-shieldings} reveal a
different picture. Here, TMHF yields the largest error with an MAE
of 24\,ppm. Generalizing the kinetic energy with the current density
only slightly reduces this error by less than 3\,ppm.
Given the poor performance for NMR shieldings, we initially suggested
that the DME approach and LMFs based on the correlation length are not well
suited for NMR properties of this test set as the corresponding functionals,
namely LHJ14, LHJ-HF, LHJ-HFcal, TM, TMh, and TMHF perform rather similarly and
do not yield good results in general. \cite{Holzer.Franzke:Local.2022}
In contrast, CHYF and CHYF-B95 significantly
improve the accuracy, which is due to the new LMF and the admixture of
exact exchange as well as the correlation term. CHYF leads to an MAE of
7.5\,ppm, which marks a similar performance as observed for r$^2$SCAN and LH20t.
The NMR-optimized functional KT3 performs best with an MAE of 6\,ppm.
Notably, the great accuracy found for the NMR shieldings with KT3 is
in strong contrast to its poor performance for NMR coupling constants.

\begin{sidewaysfigure}
\includegraphics[width=1.0\linewidth]{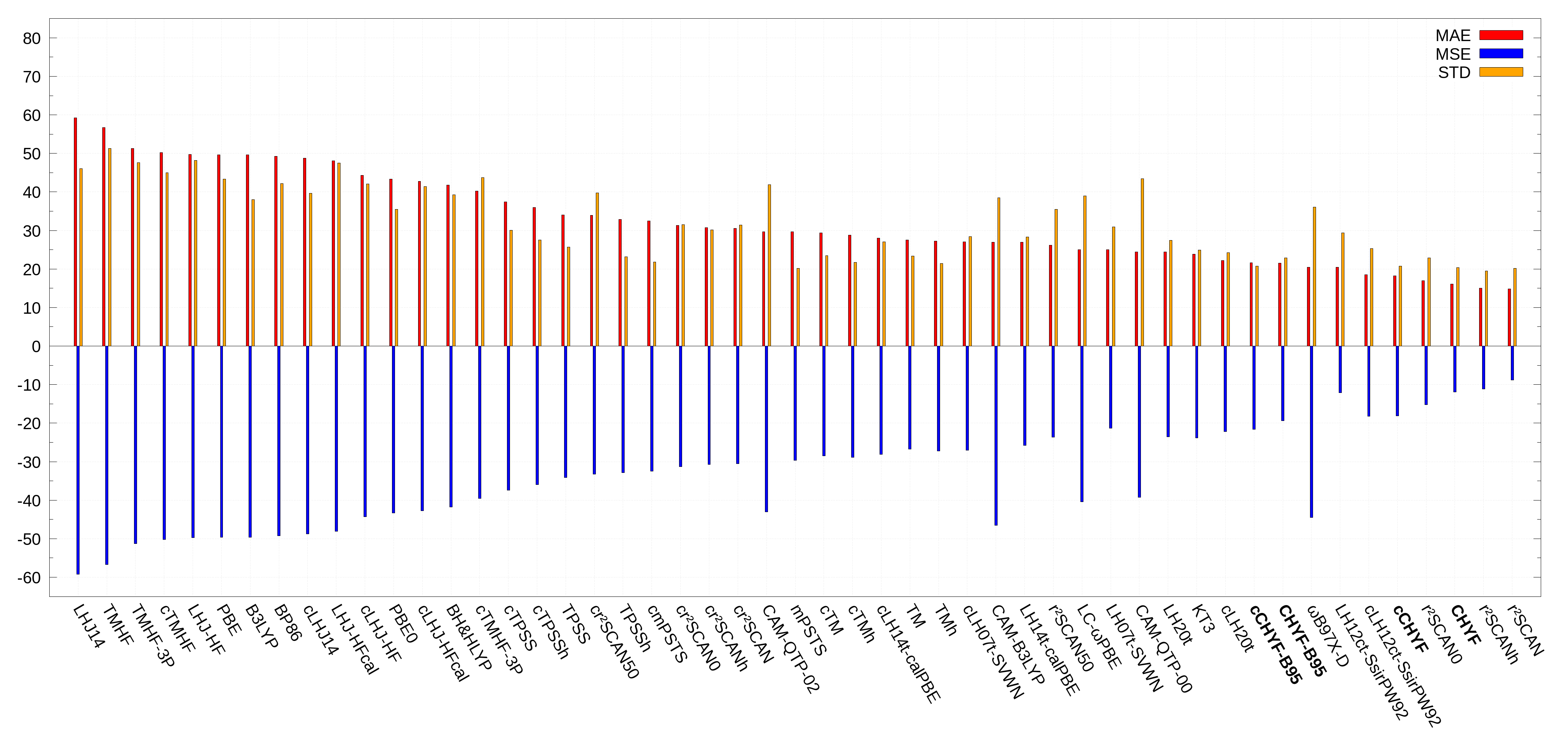}
\caption{Assessment of various density functional approximations
compared to CCSD(T) results ~\cite{Stoychev.Auer.ea:Self-Consistent.2018}
for $^{15}$N, $^{17}$O, $^{19}$F, and $^{31}$P NMR shielding constants.
MAE, MSE, and STD denote the mean absolute error, mean signed error, and its standard
deviation.}
\label{fig:nofp-shieldings}
\end{sidewaysfigure}

Results for the other shieldings in Fig.~\ref{fig:nofp-shieldings} confirm
the findings for the $^{13}$C results. Here, TMHF is one of the worst
performing functionals with an MAE of more than 50\,ppm. Only LHJ14 yields
an even larger error close to 60\,ppm. Inclusion of the current density
reduces the errors by 6 and 10\,ppm, respectively. The top performer
r$^2$SCAN leads to an MAE below 15\,ppm. CHYF reduces the MAE of TMHF
to about 16\,ppm. That is, the error is reduced by 70\%. Here,
CHYF is among the top performers, while CHYF-B95 performers somewhat
worse but errors are still acceptable.

Overall, CHYF shows an excellent performance. Therefore, CHYF
eliminates the main weakness of TMHF, as the latter leads to poor
results for this test set. Similar to many other local hybrids,
CHYF only shows a minor dependence on the generalization of
the current density for this set.

\clearpage

\clearpage

\clearpage

\clearpage
\section{$^1$H and $^{13}$C NMR Shifts of Organic Compounds}
\label{sec:shifts}

\subsection{Computational Details}
The accuracy for $^1$H and $^{13}$C NMR shifts is assessed with
the test set of Ref.~\citenum{Flaig.Maurer.ea:Benchmarking.2014}.
The def2-TZVP basis set \cite{Weigend.Ahlrichs:Balanced.2005}
and fine integration grids (grid size 4) are employed.
\cite{Treutler.Ahlrichs:Efficient.1995, Treutler:Entwicklung.1995}
Tight SCF thresholds of $10^{-9}$\,E$_{\text{h}}$ for the
ground-state energy and $10^{-9}$\,a.u.\ for the root mean square
of the change of the density matrix are applied. The response equations
for the NMR shielding calculations \cite{Haser.Ahlrichs.ea:Direct.1992,
Mas-Huniar:1999, Reiter.Mack.ea:Calculation.2018,
Schattenberg.Reiter.ea:Efficient.2020, Holzer.Franzke.ea:Assessing.2021, 
Schattenberg.Kaupp:Effect.2021, Schattenberg.Kaupp:Implementation.2021,
Holzer:improved.2020}
are converged with a threshold of $10^{-7}$\,a.u.\ for the
norm of the residuum. \cite{Furche.Krull.ea:Accelerating.2016}
Inclusion of the current density is again denoted explicitly
and the same functionals as for the NMR coupling constants are
considered. Results for other functionals than TMh, CHYF, and
CHYF-B95 are taken from Refs.~\citenum{Holzer.Franzke.ea:Assessing.2021,
Holzer.Franzke:Local.2022} and high-level coupled-cluster CCSD(T)
reference values are taken from Ref.~\citenum{Flaig.Maurer.ea:Benchmarking.2014}.

In line with previous work, \cite{Holzer.Franzke.ea:Assessing.2021,
Holzer.Franzke:Local.2022} we statistically evaluate the results for
the hydrogen, carbon, and the other nuclei. To do so,
the mean absolute error (MAE), the mean signed error (MSE),
and its standard deviation (STD) are considered.

\subsection{Results}
Results are discussed in the main text. The next pages
show graphical illustrations for completeness.

\clearpage
\begin{sidewaysfigure}
\includegraphics[width=1.0\linewidth]{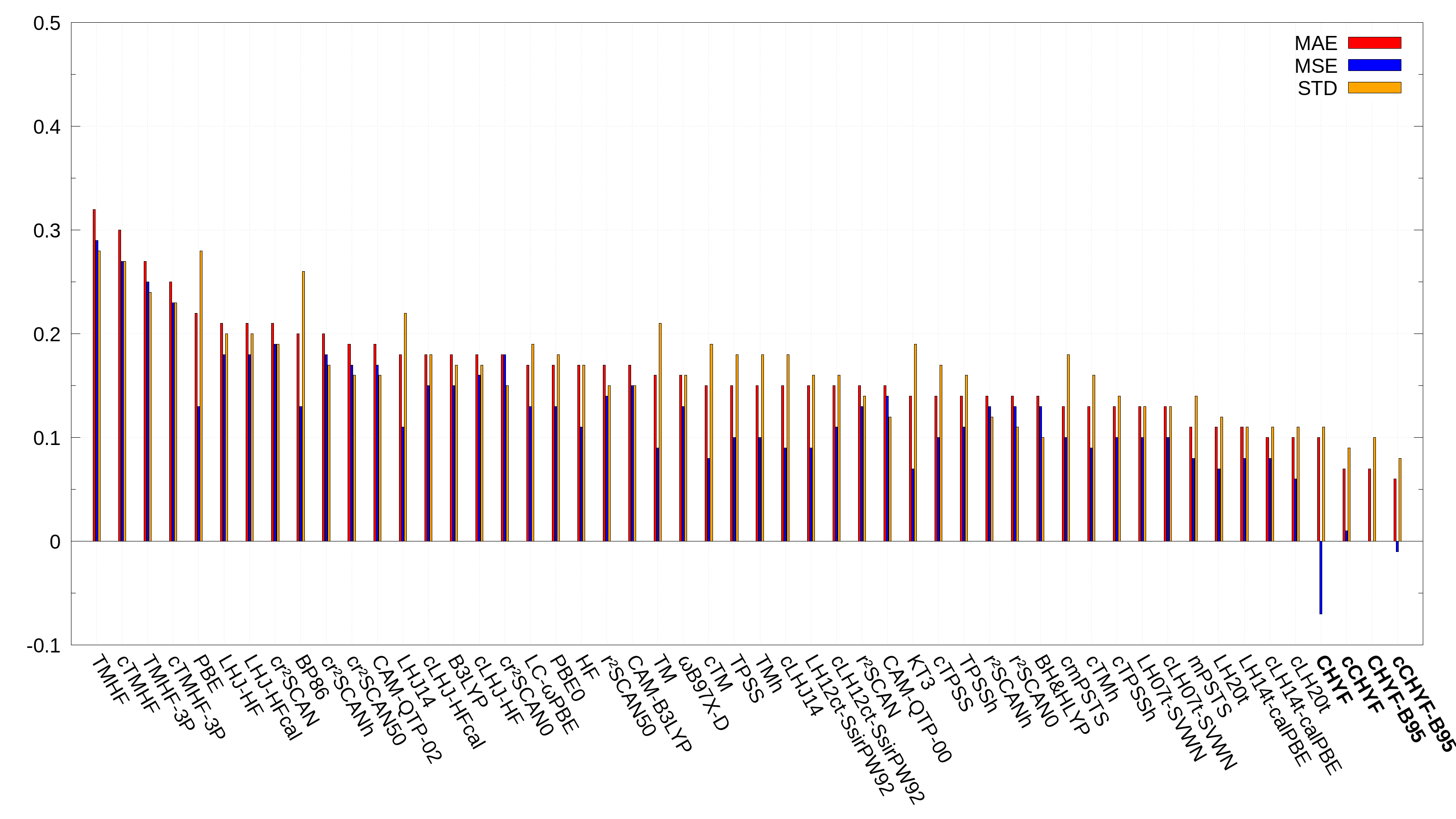}
\caption{Assessment of various density functional approximations
compared to CCSD(T) results \cite{Flaig.Maurer.ea:Benchmarking.2014}
for $^{1}$H NMR shifts.
MAE, MSE, and STD denote the mean absolute error, mean signed error,
and its standard deviation.}
\label{fig:1h-shifts}
\end{sidewaysfigure}

\clearpage
\begin{sidewaysfigure}
\includegraphics[width=1.0\linewidth]{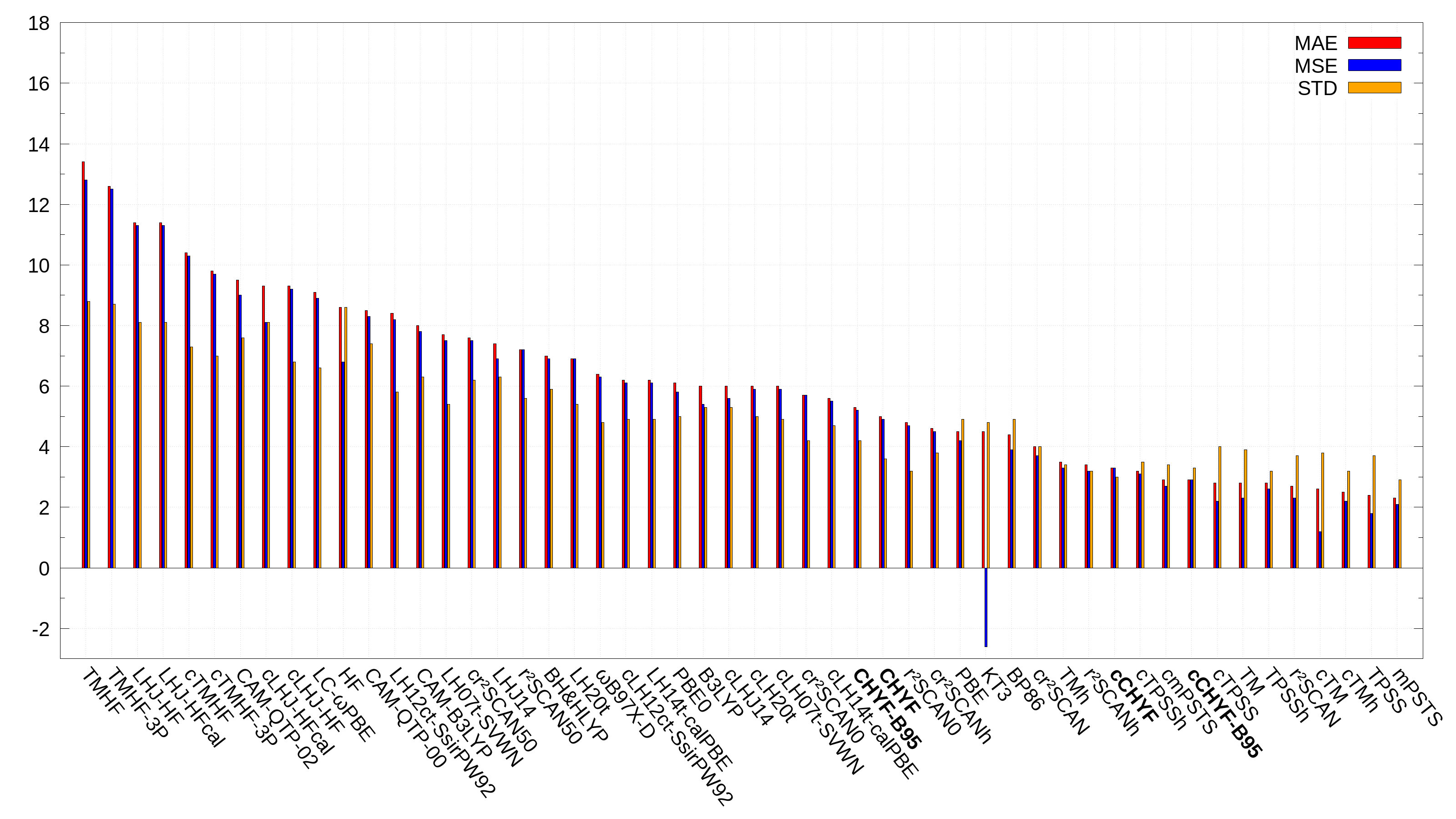}
\caption{Assessment of various density functional approximations
compared to CCSD(T) results \cite{Flaig.Maurer.ea:Benchmarking.2014}
for $^{13}$C NMR shifts.
MAE, MSE, and STD denote the mean absolute error, mean signed error,
and its standard deviation.}
\label{fig:13c-shifts}
\end{sidewaysfigure}

\clearpage
\section{Magnetizabilities}
\label{sec:magnetizabilities}

\subsection{Computational Details}
The accuracy of magnetizabilities is assessed as described in
Ref.~\citenum{Lehtola.Dimitrova.ea:Benchmarking.2021} based on
the gauge-including magnetically induced current density (GIMIC)
method. \cite{Juselius.Sundholm.ea:Calculation.2004, Taubert.Sundholm.ea:Calculation.2011,
Fliegl.Taubert.ea:gauge.2011, Sundholm.Fliegl.ea:Calculations.2016, 
Sundholm.Dimitrova.ea:Current.2021, GIMIC.2020}
Coupled-cluster CCSD(T) reference values for the test
set of 27 molecules are taken from Ref.~\citenum{Lutnas.Teale.ea:Benchmarking.2009},
which also provides the molecular structures. Note that Ozone is excluded from
the set as discussed in Ref.~\citenum{Lehtola.Dimitrova.ea:Benchmarking.2021}.
Computational settings are chosen in accordance with the literature.
\cite{Lehtola.Dimitrova.ea:Benchmarking.2021, Lehtola.Dimitrova.ea:Correction.2021, 
Lutnas.Teale.ea:Benchmarking.2009, Holzer.Franzke.ea:Assessing.2021}
The aug-cc-pCVQZ basis set \cite{Dunning:Gaussian.1989, Kendall.Dunning.ea:Electron.1992, 
Peterson.Dunning:Accurate.2002, Woon.Dunning:Gaussian.1993, Woon.Dunning:Gaussian.1995, 
Prascher.Woon.ea:Gaussian.2011} is used with large integration grids
(grid size 4) for the exchange-correlation parts. \cite{Treutler.Ahlrichs:Efficient.1995, 
Treutler:Entwicklung.1995} Tight SCF thresholds of $10^{-9}$\,E$_{\text{h}}$
for the energy and $10^{-9}$\,a.u.\ for the change of the root mean square
of the density matrix are applied. CPKS equations are converged with
a threshold of $10^{-8}$\,a.u.\ for the residuum.
Results for other functionals are taken from the literature.
\cite{Holzer.Franzke.ea:Assessing.2021} Errors are assessed
with the mean absolute error (MAE), mean signed error (MSE),
its standard deviation (STD), and the root mean square error (RMSE)

\subsection{Results}
Results are discussed in the main text. The next pages
show graphical illustrations for completeness.

\clearpage
\begin{sidewaysfigure}
\includegraphics[width=1.0\linewidth]{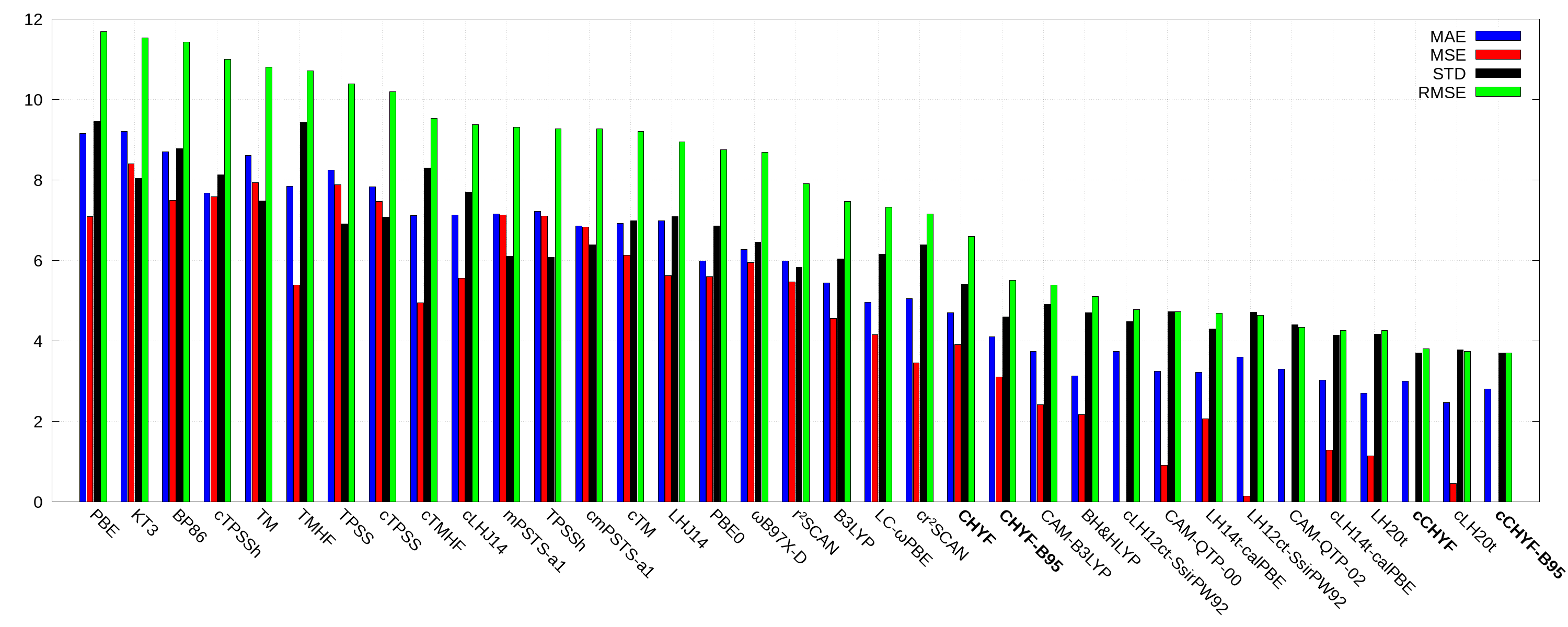}
\caption{Assessment of various density functional approximations
for magnetizabilities using the test set of Ref.~ \citenum{Lehtola.Dimitrova.ea:Benchmarking.2021}
MAE, MSE, STD, and RMSE denote the mean absolute error, mean signed
error, its standard deviation, and the root mean square error.
Errors are measured with respect to CCSD(T) results.
All values in units of $10^{-30}$ \,J/T$^2$. Functionals are sorted
according to the RMSE.}
\label{fig:magnetizability}
\end{sidewaysfigure}

\clearpage
\section{Scalar EPR Hyperfine Coupling Constants of Main-Group Systems}
\label{sec:organic-hfc}

\subsection{Computational Details}
The accuracy for the Fermi-contact (FC) term of the EPR hyperfine coupling
constant is assessed with the test sets 1 and 2 of the Bartlett group in
Ref.~\citenum{Windom.Perera.ea:Benchmarking.2022}. Here, the errors
of the DFT methods are assessed with high-level CCSD(T) and CCSD reference
values, respectively. The Fermi-contact interaction essentially probes
the density at the vicinity of the nuclei. Note that we do not
include the Be compounds of test set 1 and Zn-porphycene for test set 2,
c.f.\ Ref.\citenum{Bruder.Weigend.ea:Application.2024}.
The aug-cc-pVTZ-J basis set \cite{Provasi.Aucar.ea:effect.2001,
Provasi.Sauer:Optimized.2010, Hedegard.Kongsted.ea:Optimized.2011}
is applied and our DFT settings are the same as in 
Ref.\citenum{Bruder.Weigend.ea:Application.2024}. That is,
very large grids (grid size 5a without pruning) are employed
\cite{Franzke.Tress.ea:Error-consistent.2019} and tight thresholds
of $10^{-8}$\,E$_{\text{h}}$ for the SCF energies and $10^{-7}$\,a.u.\ for
the root mean square of the density matrix change are chosen. The
BP86, \cite{Becke:Density-functional.1988, Perdew:Density-functional.1986}
BLYP, \cite{Becke:Density-functional.1988, Lee.Yang.ea:Development.1988}
PBE, \cite{Perdew.Burke.ea:Generalized.1996}
TPSS, \cite{Tao.Perdew.ea:Climbing.2003} and
r$^2$SCAN \cite{Furness.Kaplan.ea:Accurate.2020, Furness.Kaplan.ea:Correction.2020}
PBE0, \cite{Adamo.Barone:Toward.1999}
TPSSh, \cite{Staroverov.Scuseria.ea:Comparative.2003}
r$^2$SCANh, \cite{Bursch.Neugebauer.ea:Dispersion.2022}
LC-{\textomega}PBE ,\cite{Vydrov.Scuseria:Assessment.2006}
LH12ct-SsirPW92, \cite{Arbuznikov.Kaupp:Importance.2012}
LH14t-calPBE, \cite{Arbuznikov.Kaupp:Towards.2014}
LH20t, \cite{Haasler.Maier.ea:Local.2020}
and TMHF \cite{Holzer.Franzke:Local.2022}
functionals are further applied. RPA with PBE orbitals
is included for comparison. Results for these DFT methods
are taken from Ref.\citenum{Bruder.Weigend.ea:Application.2024}.
Calculations with TM, LH12ct-SsirPW92, CHYF, and CHYF-B95 are carried out
in the present work. Errors are statistically assessed with the mean absolute
error (MAE), mean signed error (MSE), and the root mean square error (RMSE)
in MHz for the isotropic hyperfine coupling constant. Results with
LH12ct-SsirPW92 are not shown below due to the comparably large RMSE.

\clearpage
\subsection{Results}
\begin{figure}[h!]
    \centering
    \includegraphics[width=1.0\linewidth]{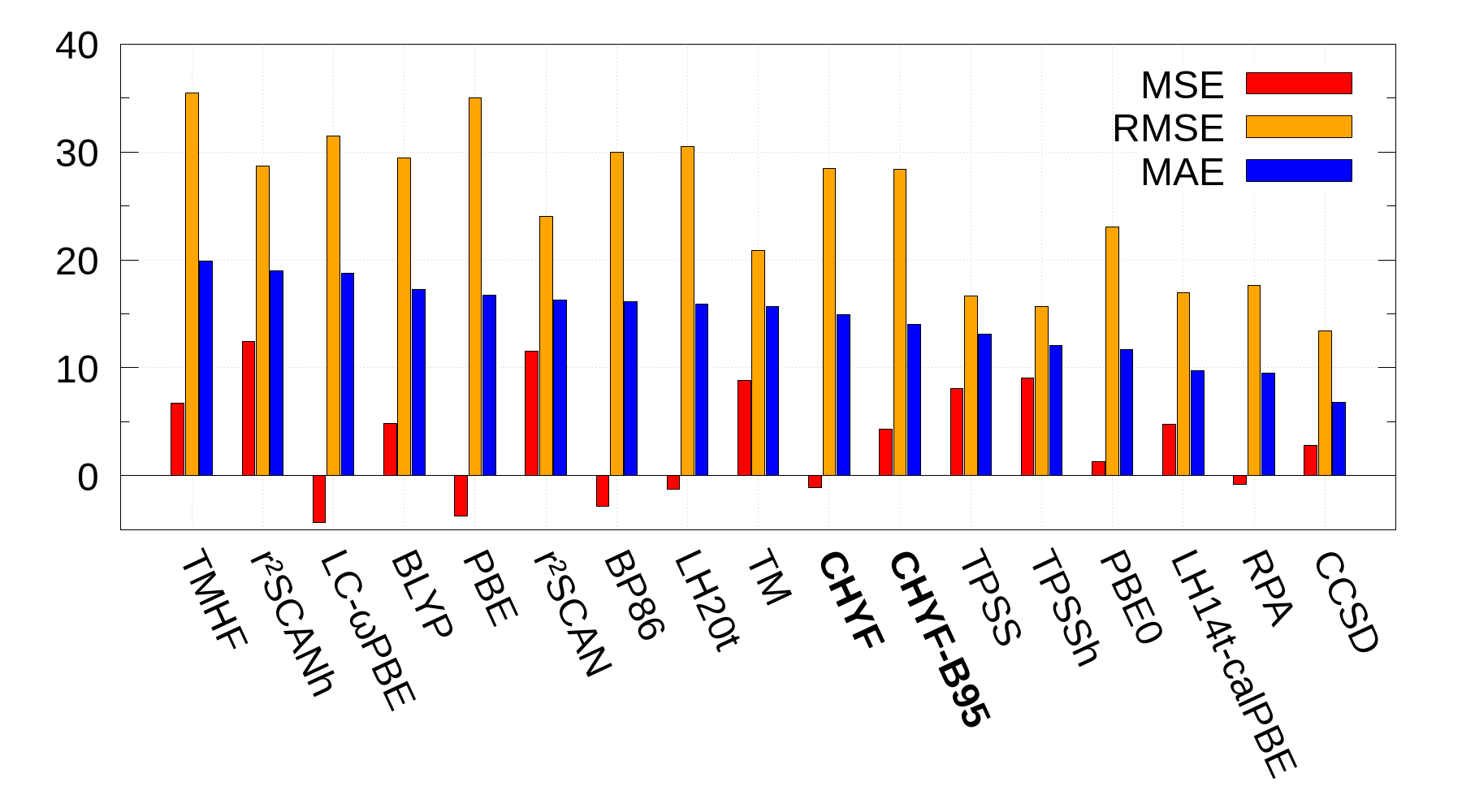}
    \caption{Statistical evaluation of DFT methods for the first test set
    Ref.~\citenum{Windom.Perera.ea:Benchmarking.2022} consisting of 23 small main-group
    radials. Deviations of the isotropic hyperfine coupling constant
    are measured with respect to CCSD(T) results in MHz.
    Data for other DFT methods than CHYF are taken from
    Ref.~\citenum{Bruder.Weigend.ea:Application.2024}. CCSD/CCSD(T)
    results are taken from Ref.~\citenum{Windom.Perera.ea:Benchmarking.2022}.
    This test set consists of twenty-two $^1$H, two $^{11}$B,
    seventeen $^{13}$C, four $^{14}$N, eight $^{17}$O, one $^{19}$F, one $^{31}$P,
    two $^{33}$S, and one $^{35}$Cl chemically inequivalent nuclei.}
    \label{fig:bartlett}
\end{figure}

Results for the first test set are depicted in Figs.~\ref{fig:bartlett}
Mean absolute errors range from about 20\,MHz for TMHF to 10\,MHz for LH14t-calPBE.
The root mean square errors cover a larger range from 35\,MHz for TMHF
to about 15\,MHz for TPSSh. Only TPSS, TPSSh, and LH14t-calPBE
yield RMSEs of less than 20\,MHz. PBE0 results in a small
MAE of 12\,MHz but its RMSE is 23\,MHz.

With an MAE of 15\,MHz CHYF performs slightly better than
the well-established local hybrid LH20t (16\,MHz)
and also outperforms its predecessor TMHF (20\,MHz).
Yet, it fails to match the accuracy of TPSSh and PBE0,
which yield MAEs of around 12\,MHz.
Considering the other functionals studied by the
Bartlett group \cite{Windom.Perera.ea:Benchmarking.2022} the
performance of CHYF and CHYF-B95 is robust and they can be safely
used for hyperfine coupling constants.

\begin{figure}[ht!]
    \centering
    \includegraphics[width=1.0\linewidth]{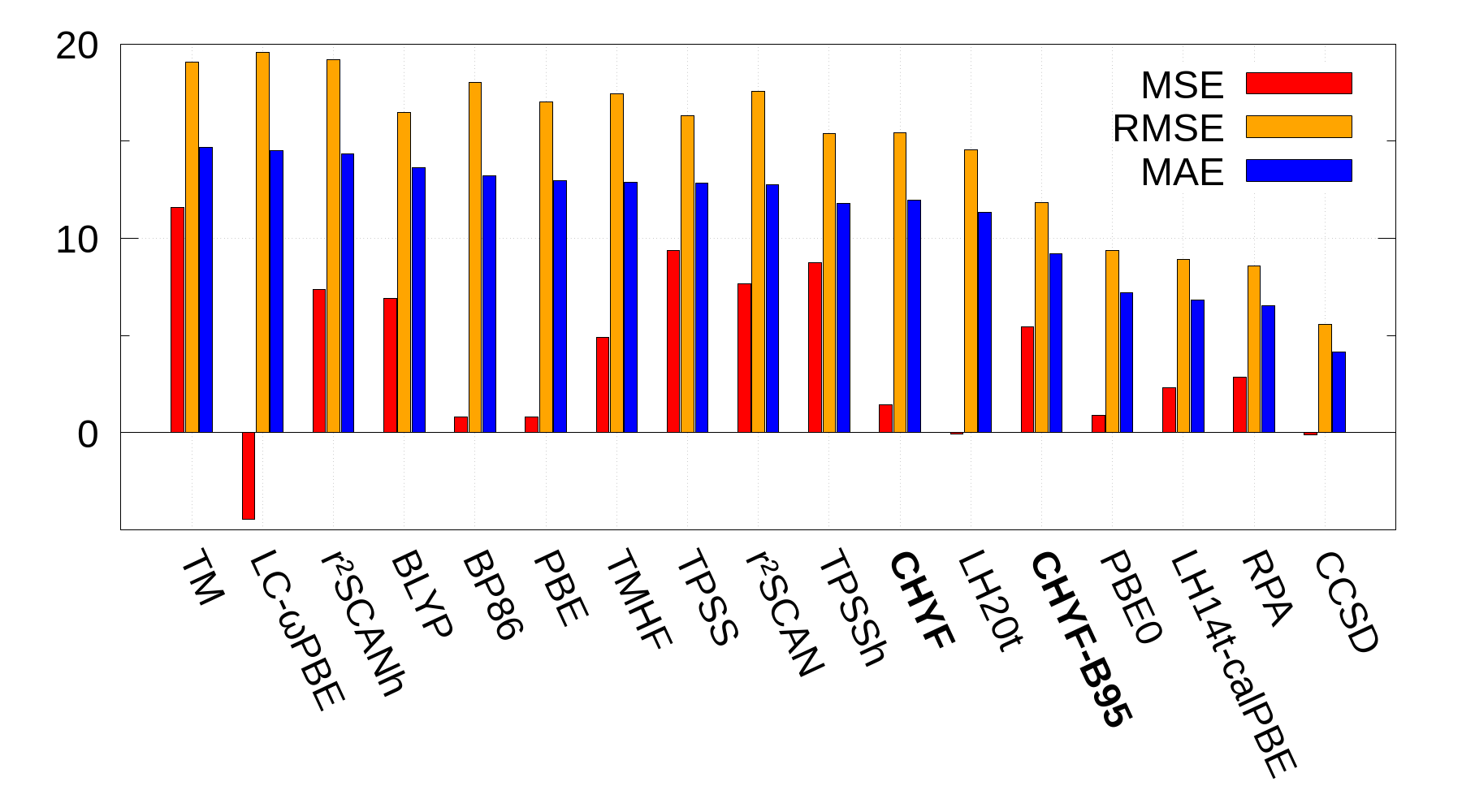}
    \caption{Statistical evaluation of DFT methods for the first test set
    Ref.~\citenum{Windom.Perera.ea:Benchmarking.2022} consisting of 23 small main-group
    radials. Compared to Fig.~\ref{fig:bartlett}, hyperfine coupling constants with
    an absolute value of more than 1000\,MHz are not considered.}
    \label{fig:bartlett-smaller}
\end{figure}

\clearpage
The hyperfine coupling constants range from about 1\,MHz to more than
1000\,MHz. This means that the very large hyperfine coupling constants
are the most important part for the statistical evaluation. To increase
the weight of the small hyperfine coupling constants, we also evaluate
the test set with hyperfine coupling constants of more than 1000\,MHz
(in absolute values). Results are shown in Fig.\ref{fig:bartlett-smaller}.
Compared to the previous findings, this significantly reduces the errors
for most functionals. Especially the RMSE is substantially reduced.
As evident from results in Fig.\ref{fig:bartlett-smaller}, PBE0
and LH14t-calPBE are the top performers with CHYF-B95, LH20t, and CHYF ranking
next. CHYF still outperforms its predecessor TMHF for all measures.
The MSE is reduced from 5\,MHz to 1\,MHz, the RMSE from 17\,MHz to 15\,MHz,
and the MAE from 13\,MHz to 11\,MHz. Overall, CHYF performs well
for the hyperfine couplings of the small main-group systems.

\clearpage
Results for the second test set reveal a somewhat different picture.
For this test set, the errors are generally smaller. All considered
functionals yield mean absolute errors between 8 and 2\,MHz.
The semilocal functionals BLYP, PBE, and BP86 show the
largest errors, whereas PBE0 and LH14t-calPBE perform best.
Based on the accuracy the functionals can be sorted into three
groups. The first one consists of BLYP, PBE, and BP86 with
comparably large errors of more than 7\,MHz.
The second group includes LH20t, r$^2$SCANh, TM, TPSS, r$^2$SCAN,
and LC-{\textomega}PBE with MAEs between 6 and 4\,MHz.
The last group is made up of the top performers CHYF-B95, CHYF,
TPSSh, TMHF, PBE0 and LH14t-calPBE with MAEs between 4 and 2\,MHz.
That is, TMHF performs well and ranks among the top functionals.
CHYF also performs very well. This shows that the accuracy for
EPR properties is rather sensitive to the test set. Nevertheless,
we can conclude that CHYF is more robust than TMHF as it works
for both NMR shifts and EPR hyperfine couplings of various test
sets.

\begin{figure}[ht!]
    \centering
    \includegraphics[width=0.9\linewidth]{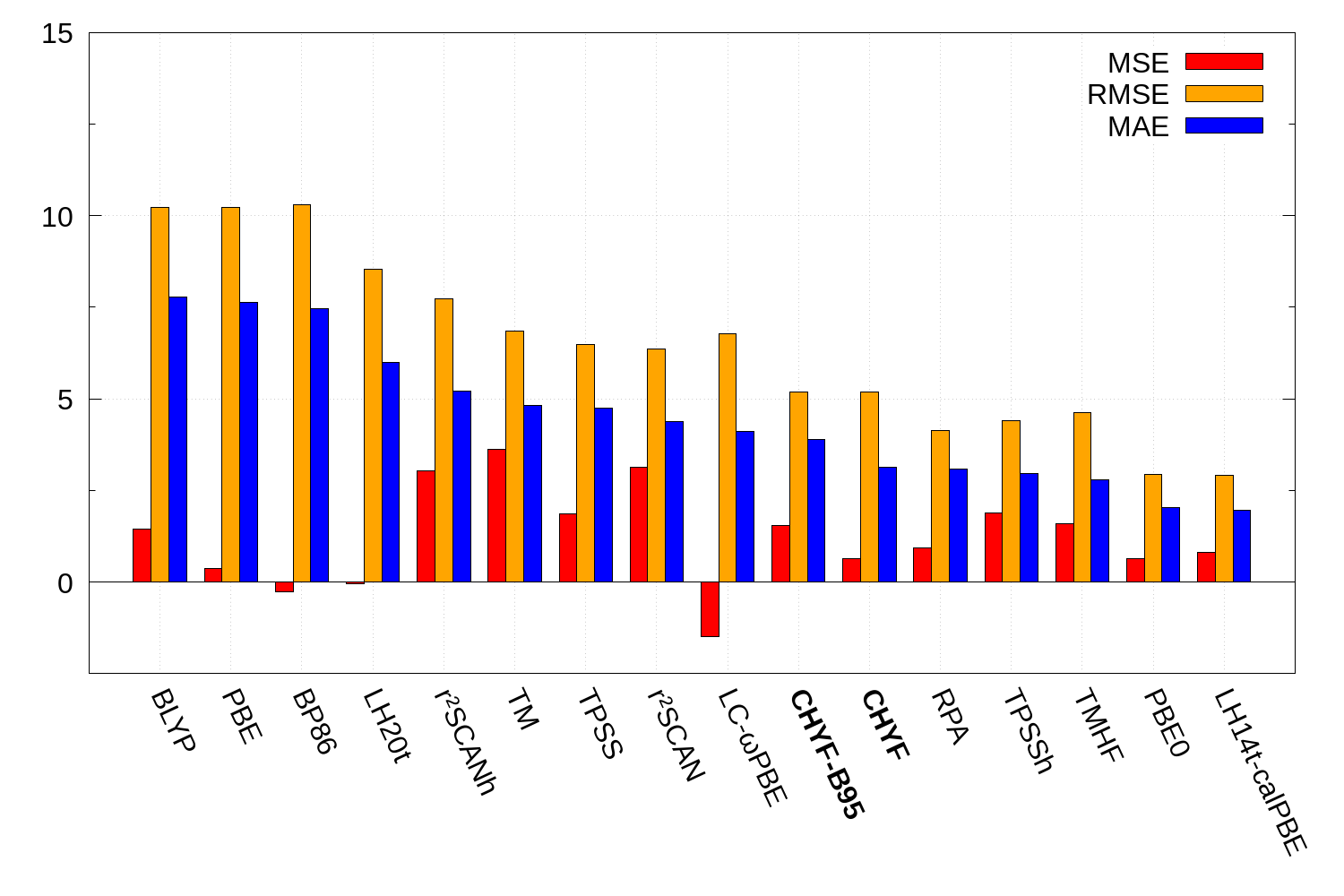}
    \caption{Statistical evaluation of various DFT methods for the second test set of 
    Ref.~\citenum{Windom.Perera.ea:Benchmarking.2022} consisting of eight larger main-group
    systems. Deviations of the isotropic hyperfine constant
    are measured with respect to CCSD results in MHz.
    Data for other DFT methods than CHYF are taken from
    Ref.~\citenum{Bruder.Weigend.ea:Application.2024}
    This test set includes thirty-three $^1$H, thirty-two $^{13}$C,
    six $^{14}$N, one $^{17}$O, and one $^{33}$S chemically different nuclei.}
    \label{fig:bartlett-2}
\end{figure}

\clearpage
\bibliography{literature}